\documentstyle[psfig]{laa}

\begin{document}

\thesaurus{11.05.1, 11.06.02, 11.16.1}

\title{
Surface Photometry of Early--type Galaxies
in the Hubble Deep Field
}

\author{
G. Fasano$^1$, M. Filippi$^2$, F. Bertola$^2$ 
}

\institute{
$^1$Astronomical Observatory of Padova, Vicolo dell' 
              Osservatorio 5, 35122 Padova, Italy \\
$^2$Department of Astronomy of the Padova University, 
 Vicolo dell' Osservatorio 5, 35122 Padova, Italy \\
e-mail gfasano@leda.pd.astro.it\\  
}

\offprints {G. Fasano}

\date{Received: September 1997.  Accepted: November 1997}

\maketitle

\markboth{Fasano et al.}{
Early--type Galaxies in the Hubble Deep Field
}

\begin{abstract}

The detailed surface photometry of a sample of early--type galaxies in
the Hubble Deep Field is presented as part of a long--term project
aimed to settle strong observational constraints to the theories
modelling the evolution of elliptical galaxies from the early stages.

The sample has been extracted, in the $V_{606}$ band, from the
database provided by the $ESO$-$STECF$-$HDF$ Group (Couch
1996). The selection criteria involve the total magnitude, the number
of pixels detected above the background level and an automatic
star/galaxy classifier. Moreover, form visual inspection of the
frames, we excluded the galaxies showing unambiguous late--type
morphology. The analysis of the luminosity and geometrical profiles,
carried out on the 162 candidates obeying our selection criteria,
resulted in a list of 99 '{\it bona fide}' early--type galaxies, for
which accurate total magnitudes and effective radii were computed on
the basis of the equivalent luminosity profiles.  The comparison with
the magnitudes given by Williams et al.(1996) indicates that the
automated photometry tends to underestimate the total luminosity of the
ellipticals.

The luminosity profiles of most of galaxies in our sample follow
fairly well the de~Vaucouleurs law (`{\it Normal}' profiles). However,
a relevant fraction of galaxies, even following the $r^{1/4}$ law in
the main body light distribution, exhibit in the inner region a
flattening of the luminosity profile not attributable to the $PSF$
(`{\it Flat}' profiles) or, in some cases, a complex (multi--nucleus)
structure (`{\it Merger}' profiles).  A statistically significant
correlation is found between the shapes of the luminosity profiles and
the ellipticity distribution. In particular, the average ellipticity
of galaxies belonging to the `{\it Flat}' and `{\it Merger}' classes
is significantly higher than that of the `{\it Normal}' galaxies.
Finally, even taken into account the relevant uncertainty of the outer
position angle profiles, the amount of isophotal twisting of HDF
ellipticals turns out to be significantly larger with respect to that
of the local samples.


\end{abstract}

\section{Introduction }

Understanding the processes governing the formation and the evolution
of galaxies from the early stages is one of the major goals of the
present day cosmology.  Comprehensive numerical codes, linked with
powerful CPU capabilities, make now possible to follow the evolution
of primordial density fluctuations down to the stages preceding the
formation of real galaxies (see Jenkins et al.\ 1997 for a recent
review). White (1997) has also shown that the scaling relations
expected after evolution of proto--galaxies of different morphologies
seem to be in good agreement with those observed today, i.e. the
Tully-Fisher (1977) and the Fundamental Plane (Dressler et al.\ 1987;
Djorgovski \& Davis 1987) relations.

Still, a number of key questions remain to be answered. We mention
(among the others) the following ones: which is the parameter driving
galaxies towards different morphologies? Are elliptical galaxies
originated from gravitational collapse of primordial fluctuations
(single burst of star formation whose duration depends on the onset of
galactic winds) or are they the result of multiple merging (infall and
recursive bursts)? Which is the influence of the internal and
intergalactic absoption in determining the observed brightness and
color profiles of high redshift galaxies?

Besides the global approach to the star formation history (Madau et
al.\ 1997), answering these questions also requires detailed
morphological and dynamical studies of galaxies in the early evolutionary
stages. In particular, luminosity and geometrical profiles in
different bands are needed to obtain unambiguous morphological
classifications as well as reliable estimates of the galaxy sizes at
intermediate and high redshifts.

Recent advances in the sensitivity and resolution of the observations
both in imaging and spectroscopy with the {\sl Hubble Space Telescope}
(HST) and from the ground have greatly enlarged the horizon of
morphological and dynamical studies for high redshift galaxies 
(Giavalisco et al.\ 1996, van~Dokkum and Franx 1996, Shade et al.\ 1996, 
Shade et al.\ 1997, Oemler et al.\ 1997, Lowenthal et al.\ 1997,
Pettini et al.\ 1997 and references therein). The Hubble Deep Field
(HDF) is perhaps the most impressive example of these progresses
(Ellis 1997).

The HDF project (Williams et al. 1996) has been realized using
observational procedures ({\it dithering}) and data handling
techniques ({\it drizzling}) aimed to improve not only the cosmetic of
the final images, but also the resolution performances. In particular,
after '{\it drizzling}', the pixel size in arcseconds of the three WF
cameras turns out to be even better than that of the PC
($\sim0^{\prime\prime}.040$ vs. $\sim0^{\prime\prime}.046$). Moreover,
the very long total exposure times of the final images allow to
overcome the main limitation of the WFPC data, that is the relatively
high surface brightness level usually reached for extended
sources. This makes the HDF frames particularly suited in order to
perform the detailed surface photometry of objects whose surface
luminosity slowly decreases outwards down to very faint levels, as in
the case of elliptical galaxies. We can rightfully assert that the HDF
represents the best opportunity we had since now to study the
morphology of elliptical galaxies at very high redshifts.

We have undertaken a long--term project aimed to produce the detailed
surface photometry, in different bands, for a sample of early--type
galaxies in the HDF. The main scientific goal of this project is to
settle strong observational constraints to the theories modelling the
evolution of elliptical galaxies from the early stages (e.g. Tantalo
et al.\ 1996, Kauffmann and Charlot 1997, Chiosi et al.\ 1997 and 
references therein).
 
In this paper we present the surface photometry in the $V_{606}$
band. In the forthcoming papers (Filippi et al. 1997, Fasano et al. 1997)
we will present the surface photometry in the remaining 3 bands and
discuss the improvements that the morphological information,
together with the photometry in different optical and infrared bands,
can produce in understanding the processes of galaxy formation and
evolution.

In Section 2 we discuss the sample selection. Section 3 illustrates
the techniques we used to extract the morphological information from
the HDF frames. In Section 4 we present the results of the detailed
surface photometry and discuss the global morphological properties of
the sample. 

\section{
Sample selection
}

Our sample of {\it early--type} galaxies has been extracted from the
second release of the WFPC2--HDF frames, in the $V_{606}$ band. Due
to the efficiency curve of the WFPC2, this filter usually ensures a
better S/N ratio with respect to the other available filters
($U_{300}$, $B_{450}$ and $I_{814}$) even in the case of high redshift
galaxies, as we verified by comparing the $V_{606}$ and $I_{814}$ images.

\subsection{Selection criteria}

The sample selection is based on the compilation provided by the
$ESO$-$STECF$-$HDF$ Group, obtained through the automated SExtractor
algorithm (Bertin and Arnouts 1996). Among the other things, the
catalog includes, for each object in the field, the total $V_{606}$
magnitude in the STMAG system, the number of pixels ($N_{pix}$)
detected above the background using a threshold of $1.3\sigma$ of the
noise, and a star/galaxy $(s/g)$ classifier, ranging from 0 to 1, which
gives the probability that the object is stellar ($s/g$=1 means
'star').  After a preliminar inspection of the frames, supplemented by
several fast average--tracing luminosity profiles of faint and/or
small objects, we decided to set the following limits for inclusion in
our preliminar sample:
(1) $V_{606}(STMAG)\ \le$ 26.5;
(2) $N_{pix}\ \ge$ 200;
(3) $(s/g)\ \le$ 0.6.

As for the first two limits, they concern essentially the technical
'{\it feasibility}' of the morphological analysis. Although at this
stage we preferred not to be too much severe in the selection, these
limits provide completeness criteria to our sample. The choice
of the third limit turned out not to be critical, probably due to the
very shape of the star profiles. Nevertheless, in this case we
preferred to be more conservative, in order to prevent the inclusion
of stars in our sample.

\subsection{Morphological screening}

We found 372 objects of the SExtractor catalog matching the above
limits in the three chips of the WFC (the catalog does not include the
PC). 29 objects from the chip $\#1$ (PC) were added later to
this list by simple visual inspection, so that the total preliminar
sample turned out to be of 401 objects.

We used the '{\it imexam}' IRAF's tool, together with the '{\it
Saoimage}' display facility, to estimate the
morphological type of each object, thus producing a first screening of
the galaxies in the sample. Apart from some cases, the angular
resolution did not allow us to give precise Hubble
types. Moreover, many faint galaxies appear heavily disturbed by the
presence of close (often multiple) companions or peculiar structures,
which obviously make even more difficult the classification.
Nevertheless, most of the '{\it early--type}' candidates turned out to
be easily recognizable on the basis of the light concentration. The
doubious cases (mainly $S0s$ and $Sa$ galaxies) were retained at this
stage and were examined later in a quantitative way (see next Section).

After this selection we were left with 134 galaxies. In order to check 
the reliability of our
classifications, we compared them with those given by Van den Bergh
et al.(1996) and by Statler(1996, private communication). Due to
the different selection criteria, the three samples do not overlap 
exactly each other. Nevertheless, for the common objects the agreement 
was fairly good ($80\%$ with Van den Bergh et al. and $84\%$ with Statler). 
To be more
accurate, we decided to provisionally include in our sample all galaxies
matching our selection criteria which, contrary to our estimates, were 
classified as '{\it early--type}' by Van den Bergh et al.(1996) and/or 
by Statler(1996). In this way 28 more galaxies enriched our sample, which
become of 162 objects. 

\subsection{Final sample}

We performed the detailed surface photometry of this sample,
producing luminosity and geometrical profiles of each object. The
next Section outlines the techniques we used to extract the
morphological information from this very peculiar observational
material. From the analysis of the luminosity and geometrical
profiles, 34 objects of the sample have been recognized to be '{\it
disk--dominated}' objects (likely $Sa$ galaxies), whereas 28 more objects
showed peculiar or unclassifiable profiles. These galaxies were
excluded from the final sample of '{\it bona fide}' early--type
galaxies. 

The final sample of 99 galaxies is reported
in Table 1, where column (1) gives the FOCAS list identification
(Williams et al. 1996), columns (2) and (3) the J2000 coordinates
$\alpha$ and $\delta$, column (4) our morphological classification,
columns (5) the morphological type (if available) given by
van den Bergh et al.(1996). 

\begin{table*}
\caption[ ]{ The Sample}
\begin{tabular}{l|ll|cc|c|cc|ccr}
\hline
\hline
FOCAS--ID          &\multicolumn{2}{c|}{Coordinates(2000)}  & \multicolumn{2}{c|} 
{Morphology} &Lum.Prof. &$V_{606}^{Tot}(\Delta_m)$ &$r_e^{eq}$ & $\varepsilon_e$ & $\varepsilon_{max}$ & $\Delta\theta$ \\ 
& \multicolumn{1}{c}{$\alpha$} & \multicolumn{1}{c|}{$\delta$} &Our &V.d.Bergh &class & ($STMAG$) & ($^{\prime\prime}$) & & & ($^\circ$) \\
\hline
ID\_4\_942\_0    & $36^m~39^s.43$ & $12^{\prime}~11^{\prime\prime}.76$ & E    & E   & 1 & 24.75(.33) & 0.54 & 0.18 & 0.25 &  15 \\ 
ID\_4\_926\_2    & $36\ \ \ 39.56$ & $12\ \ 13.83$ & m    & p    & 3 & 25.36(.17) & 0.19 & 0.04 & 0.17 &  60 \\
ID\_4\_928\_1    & $36\ \ \ 40.01$ & $12\ \ 07.37$ & S0   & E    & 1 & 23.61(.15) & 0.36 & 0.10 & 0.19 &  50 \\
ID\_4\_878\_2    & $36\ \ \ 40.74$ & $12\ \ 04.96$ & E    & E    & 2 & 25.79(.19) & 0.22 & 0.35 & 0.35 &   8 \\
ID\_4\_878\_11   & $36\ \ \ 40.96$ & $12\ \ 05.31$ & E    & Sap  & 2 & 23.66(.14) & 0.22 & 0.12 & 0.14 &  30 \\
ID\_4\_822\_0    & $36\ \ \ 41.15$ & $12\ \ 10.56$ & E    & E    & 1 & 25.42(.27) & 0.33 & 0.08 & 0.10 &   5 \\
ID\_4\_858\_13   & $36\ \ \ 41.25$ & $12\ \ 03.07$ & E    & E    & 1 & 25.18(.23) & 0.15 & 0.14 & 0.19 &  11 \\
ID\_4\_767\_0    & $36\ \ \ 41.49$ & $12\ \ 14.98$ & S0   & E    & 2 & 25.16(.15) & 0.19 & 0.36 & 0.39 &   6 \\
ID\_4\_661\_1    & $36\ \ \ 41.62$ & $12\ \ 35.67$ & E    & E    & 2 & 24.96(.17) & 0.22 & 0.18 & 0.23 &  20 \\
ID\_4\_639\_1    & $36\ \ \ 41.71$ & $12\ \ 38.75$ & E    & Et?  & 1 & 25.12(.14) & 0.17 & 0.08 & 0.16 &  47 \\
ID\_4\_937\_0    & $36\ \ \ 42.28$ & $11\ \ 26.18$ & E    & E    & 1 & 25.50(.19) & 0.18 & 0.19 & 0.22 &  30 \\
ID\_1\_95\_0     & $36\ \ \ 42.38$ & $13\ \ 19.36$ & E    & --   & 1 & 25.03(.13) & 0.13 & 0.28 & 0.30 &  12 \\
ID\_4\_602\_0    & $36\ \ \ 42.41$ & $12\ \ 32.48$ & E    & E    & 1 & 25.30(.18) & 0.16 & 0.24 & 0.28 &  15 \\
ID\_4\_804\_0    & $36\ \ \ 42.53$ & $11\ \ 50.01$ & E    & Ep?  & 1 & 25.40(.16) & 0.20 & 0.23 & 0.27 &   8 \\
ID\_4\_774\_3    & $36\ \ \ 42.78$ & $11\ \ 54.28$ & E    & --   & 1 & 25.95(.18) & 0.14 & 0.20 & 0.21 &  26 \\
ID\_4\_581\_11   & $36\ \ \ 42.87$ & $12\ \ 27.85$ & E/S0 & --   & 1 & 25.77(.15) & 0.14 & 0.19 & 0.25 &  40 \\
ID\_4\_845\_0    & $36\ \ \ 42.92$ & $11\ \ 37.27$ & E    & Sa   & 2 & 25.08(.16) & 0.20 & 0.37 & 0.37 &   3 \\
ID\_4\_554\_1    & $36\ \ \ 43.13$ & $12\ \ 28.11$ & Ep   & Ep   & 2 & 25.58(.20) & 0.17 & 0.29 & 0.33 &  38 \\
ID\_4\_493\_0    & $36\ \ \ 43.16$ & $12\ \ 42.20$ & S0   & E    & 1 & 23.26(.24) & 0.44 & 0.33 & 0.33 &   7 \\
ID\_4\_727\_0    & $36\ \ \ 43.41$ & $11\ \ 51.57$ & E    & Ep?  & 2 & 23.43(.10) & 0.21 & 0.49 & 0.48 &   3 \\
ID\_4\_565\_0    & $36\ \ \ 43.63$ & $12\ \ 18.25$ & m    & Sap  & 3 & 23.46(.12) & 0.31 & 0.18 & 0.28 &  65 \\
ID\_4\_744\_0    & $36\ \ \ 43.80$ & $11\ \ 42.88$ & E    & --   & 1 & 22.33(.18) & 0.63 & 0.12 & 0.16 &  50 \\
ID\_2\_82\_1     & $36\ \ \ 44.07$ & $14\ \ 09.92$ & E    & E/m  & 2 & 24.92(.15) & 0.18 & 0.27 & 0.28 &  19 \\
ID\_4\_752\_1    & $36\ \ \ 44.38$ & $11\ \ 33.20$ & E    & --   & 1 & 22.97(.22) & 0.90 & 0.18 & 0.18 &   3 \\
ID\_1\_56\_0     & $36\ \ \ 44.46$ & $13\ \ 13.10$ & E    & --   & 1 & 26.38(.19) & 0.13 & 0.15 & 0.21 &   7 \\
ID\_4\_579\_0    & $36\ \ \ 44.74$ & $11\ \ 57.05$ & E    & E/*  & 1 & 25.53(.14) & 0.13 & 0.06 & 0.09 &  93 \\
ID\_1\_37\_1     & $36\ \ \ 44.79$ & $13\ \ 07.23$ & m    & --   & 3 & 25.75(.18) & 0.13 & 0.50 & 0.46 &  28 \\
ID\_2\_163\_0    & $36\ \ \ 45.29$ & $14\ \ 07.03$ & E    & --   & 1 & 25.94(.19) & 0.18 & 0.25 & 0.26 &  15 \\
ID\_4\_555\_2    & $36\ \ \ 45.33$ & $11\ \ 54.52$ & E/S0 & E    & 1 & 24.79(.21) & 0.32 & 0.17 & 0.24 &  20 \\
ID\_4\_368\_0    & $36\ \ \ 45.35$ & $12\ \ 33.70$ & E/S0 & --   & 1 & 26.07(.22) & 0.24 & 0.26 & 0.25 &  25 \\
ID\_2\_80\_0     & $36\ \ \ 45.40$ & $13\ \ 50.07$ & E    & --   & 2 & 25.64(.18) & 0.18 & 0.26 & 0.28 &   8 \\
ID\_1\_100\_0    & $36\ \ \ 45.52$ & $13\ \ 29.97$ & E    & --   & 1 & 26.37(.14) & 0.13 & 0.12 & 0.29 &  22 \\
ID\_1\_35\_0     & $36\ \ \ 45.61$ & $13\ \ 08.92$ & E    & --   & 1 & 24.35(.16) & 0.21 & 0.24 & 0.26 &  14 \\
ID\_4\_516\_0    & $36\ \ \ 45.65$ & $11\ \ 53.97$ & E    & Et   & 1 & 25.30(.15) & 0.18 & 0.13 & 0.20 &  40 \\
ID\_4\_497\_0    & $36\ \ \ 45.73$ & $11\ \ 57.31$ & E    & --   & 1 & 26.05(.17) & 0.13 & 0.12 & 0.14 &  20 \\
ID\_4\_520\_0    & $36\ \ \ 45.79$ & $11\ \ 50.52$ & E/S0 & --   & 1 & 26.07(.19) & 0.14 & 0.11 & 0.18 &  15 \\
ID\_2\_61\_0     & $36\ \ \ 46.12$ & $13\ \ 34.62$ & E    & E    & 2 & 25.91(.15) & 0.15 & 0.26 & 0.31 &  10 \\
ID\_4\_254\_0    & $36\ \ \ 46.13$ & $12\ \ 46.50$ & E    & E/*  & 1 & 23.64(.19) & 0.55 & 0.27 & 0.17 &  45 \\
ID\_1\_47\_0     & $36\ \ \ 46.16$ & $13\ \ 13.89$ & E    & --   & 2 & 25.17(.13) & 0.15 & 0.38 & 0.37 &  16 \\
ID\_4\_322\_2    & $36\ \ \ 46.21$ & $12\ \ 28.43$ & E    & Et   & 2 & 25.59(.16) & 0.16 & 0.23 & 0.28 &   5 \\
ID\_2\_251\_0    & $36\ \ \ 46.34$ & $14\ \ 04.62$ & E/S0 & --   & 1 & 22.96(.17) & 0.52 & 0.05 & 0.14 & 110 \\
ID\_4\_471\_0    & $36\ \ \ 46.51$ & $11\ \ 51.32$ & E    & E    & 1 & 23.11(.07) & 0.22 & 0.08 & 0.12 &  45 \\
ID\_4\_289\_0    & $36\ \ \ 46.95$ & $12\ \ 26.08$ & E    & E    & 1 & 25.73(.14) & 0.16 & 0.11 & 0.20 &  49 \\
ID\_2\_201\_0    & $36\ \ \ 47.18$ & $13\ \ 41.82$ & E    & Et   & 1 & 24.45(.16) & 0.14 & 0.15 & 0.24 &  15 \\
ID\_2\_272\_0    & $36\ \ \ 47.68$ & $13\ \ 51.28$ & E    & --   & 2 & 25.98(.16) & 0.17 & 0.26 & 0.28 &  34 \\
ID\_2\_363\_0    & $36\ \ \ 47.71$ & $14\ \ 09.43$ & E    & --   & 2 & 25.94(.15) & 0.15 & 0.25 & 0.33 &  17 \\
ID\_2\_121\_111  & $36\ \ \ 48.08$ & $13\ \ 09.02$ & S0   & --   & 1 & 21.48(.11) & 0.60 & 0.04 & 0.23 &  30 \\
ID\_2\_412\_11   & $36\ \ \ 48.11$ & $14\ \ 14.42$ & E/S0 & Sap  & 2 & 25.42(.22) & 0.17 & 0.27 & 0.42 &   7 \\
ID\_4\_260\_112  & $36\ \ \ 48.12$ & $12\ \ 14.90$ & E    & E    & 2 & 25.08(.13) & 0.18 & 0.30 & 0.28 &  13 \\
ID\_2\_449\_1    & $36\ \ \ 48.34$ & $14\ \ 16.63$ & E    & E    & 2 & 24.11(.22) & 0.22 & 0.24 & 0.30 &  15 \\
ID\_2\_173\_0    & $36\ \ \ 48.47$ & $13\ \ 16.62$ & E    & E    & 2 & 23.74(.29) & 0.72 & 0.18 & 0.35 &   8 \\
ID\_2\_537\_12   & $36\ \ \ 48.71$ & $14\ \ 22.62$ & E    & Et   & 2 & 25.60(.15) & 0.20 & 0.32 & 0.35 &  12 \\
ID\_3\_51\_0     & $36\ \ \ 48.72$ & $13\ \ 02.45$ & E    & E/*  & 1 & 25.93(.17) & 0.17 & 0.08 & 0.14 &   5 \\
ID\_2\_236\_2    & $36\ \ \ 48.97$ & $13\ \ 21.88$ & E    & Et   & 1 & 24.81(.10) & 0.16 & 0.03 & 0.13 &  65 \\
ID\_2\_180\_0    & $36\ \ \ 49.05$ & $13\ \ 09.64$ & E    & Sa   & 2 & 24.65(.15) & 0.20 & 0.50 & 0.47 &   4 \\
ID\_4\_274\_0    & $36\ \ \ 49.11$ & $11\ \ 50.54$ & S0   & --   & 2 & 26.02(.25) & 0.17 & 0.23 & 0.24 &  10 \\
ID\_2\_264\_2    & $36\ \ \ 49.38$ & $13\ \ 11.22$ & E    & E    & 1 & 22.80(.11) & 0.21 & 0.24 & 0.26 &  30 \\
ID\_2\_456\_1111 & $36\ \ \ 49.44$ & $13\ \ 46.88$ & S0   & --   & 1 & 18.93(.05) & 0.52 & 0.37 & 0.39 &   7 \\
ID\_3\_229\_0    & $36\ \ \ 49.48$ & $12\ \ 48.73$ & E    & E/*  & 2 & 24.96(.24) & 0.24 & 0.10 & 0.12 &  15 \\
ID\_2\_590\_1    & $36\ \ \ 49.51$ & $14\ \ 21.10$ & E    & --   & 2 & 25.89(.21) & 0.16 & 0.18 & 0.16 &  55 \\
ID\_4\_109\_0    & $36\ \ \ 49.60$ & $12\ \ 12.68$ & E    & p    & 2 & 25.29(.12) & 0.16 & 0.38 & 0.40 &  10 \\
ID\_3\_143\_0    & $36\ \ \ 49.64$ & $12\ \ 57.43$ & m    & Sap  & 3 & 22.85(.15) & 0.32 & 0.55 & 0.65 &   4 \\
\hline
\end{tabular}
\end{table*}
\setcounter{table}{0}
\begin{table*}
\caption[ ]{...continue...}
\begin{tabular}{l|ll|cc|c|cc|ccr}
\hline
\hline
FOCAS--ID          &\multicolumn{2}{c|}{Coordinates(2000)}  & \multicolumn{2}{c|} 
{Morphology} &Lum.Prof. &$V_{606}^{Tot}(\Delta_m)$ &$r_e^{eq}$ & $\varepsilon_e$ & $\varepsilon_{max}$ & $\Delta\theta$ \\ 
& \multicolumn{1}{c}{$\alpha$} & \multicolumn{1}{c|}{$\delta$} &Our &V.d.Bergh &class & ($STMAG$) & ($^{\prime\prime}$) & & & ($^\circ$) \\
\hline
ID\_3\_243\_0    & $36\ \ \ 49.81$ & $12\ \ 48.79$ & E/S0p& Ep   & 1 & 25.52(.21) & 0.21 & 0.32 & 0.34 &  14 \\
ID\_2\_456\_22   & $36\ \ \ 50.03$ & $13\ \ 51.99$ & S0   & Ep   & 1 & 25.15(.12) & 0.17 & 0.36 & 0.50 &   6 \\
ID\_3\_321\_1    & $36\ \ \ 50.27$ & $12\ \ 45.75$ & S0   & --   & 1 & 23.05(.13) & 0.43 & 0.34 & 0.36 &   5 \\
ID\_2\_373\_0    & $36\ \ \ 50.30$ & $13\ \ 29.73$ & E    & E    & 1 & 25.58(.18) & 0.16 & 0.27 & 0.33 &   9 \\
ID\_2\_725\_0    & $36\ \ \ 50.56$ & $14\ \ 28.47$ & E    & E    & 2 & 25.50(.15) & 0.17 & 0.52 & 0.44 &  12 \\
ID\_2\_693\_1    & $36\ \ \ 51.35$ & $14\ \ 11.02$ & E    & E    & 2 & 25.22(.15) & 0.18 & 0.18 & 0.25 &  22 \\
ID\_3\_659\_2    & $36\ \ \ 51.44$ & $12\ \ 20.71$ & Ep   & E/S0 & 1 & 25.69(.15) & 0.13 & 0.18 & 0.17 &  23 \\
ID\_3\_902\_1    & $36\ \ \ 51.79$ & $11\ \ 57.81$ & E/Sa & --   & 2 & 26.25(.22) & 0.16 & 0.18 & 0.20 &  42 \\
ID\_2\_531\_0    & $36\ \ \ 51.97$ & $13\ \ 32.18$ & E    & E    & 3 & 24.06(.15) & 0.29 & 0.10 & 0.07 &  15 \\
ID\_3\_586\_0    & $36\ \ \ 52.10$ & $12\ \ 26.31$ & E    & E/Sa & 1 & 25.77(.36) & 0.36 & 0.14 & 0.19 &  20 \\
ID\_2\_646\_0    & $36\ \ \ 52.23$ & $13\ \ 48.07$ & E    & E    & 2 & 25.14(.21) & 0.20 & 0.42 & 0.38 &   5 \\
ID\_2\_849\_0    & $36\ \ \ 52.40$ & $14\ \ 20.95$ & E    & Ep   & 2 & 25.60(.26) & 0.19 & 0.32 & 0.30 &  15 \\
ID\_3\_625\_0    & $36\ \ \ 52.51$ & $12\ \ 24.78$ & m    & p    & 3 & 25.39(.13) & 0.15 & 0.50 & 0.49 &   3 \\
ID\_3\_696\_0    & $36\ \ \ 52.69$ & $12\ \ 19.72$ & E    & E    & 2 & 23.77(.09) & 0.19 & 0.46 & 0.45 &  10 \\
ID\_2\_637\_0    & $36\ \ \ 52.76$ & $13\ \ 39.08$ & E    & E/*  & 1 & 25.18(.15) & 0.18 & 0.11 & 0.14 &  80 \\
ID\_3\_886\_0    & $36\ \ \ 52.92$ & $12\ \ 03.11$ & E    & E    & 2 & 25.21(.18) & 0.14 & 0.16 & 0.21 &   8 \\
ID\_2\_726\_1    & $36\ \ \ 53.12$ & $13\ \ 46.25$ & E    & --   & 1 & 26.75(.19) & 0.11 & 0.23 & 0.18 &  35 \\
ID\_2\_635\_0    & $36\ \ \ 53.15$ & $13\ \ 31.66$ & E    & --   & 1 & 25.99(.15) & 0.17 & 0.24 & 0.24 &  29 \\
ID\_2\_591\_2    & $36\ \ \ 53.18$ & $13\ \ 22.75$ & E    & --   & 2 & 25.05(.15) & 0.18 & 0.31 & 0.30 &  10 \\
ID\_3\_670\_1    & $36\ \ \ 53.26$ & $12\ \ 22.74$ & E    & E    & 2 & 25.73(.21) & 0.14 & 0.28 & 0.26 &  20 \\
ID\_2\_643\_0    & $36\ \ \ 53.42$ & $13\ \ 29.52$ & m    & E+E  & 3 & 25.03(.17) & 0.22 & 0.65 & 0.62 &   5 \\
ID\_2\_973\_2    & $36\ \ \ 54.50$ & $14\ \ 08.16$ & E    & --   & 2 & 25.88(.18) & 0.18 & 0.23 & 0.19 &  10 \\
ID\_3\_118\_1    & $36\ \ \ 54.73$ & $13\ \ 14.73$ & E    & E/*  & 1 & 24.59(.17) & 0.14 & 0.08 & 0.09 &  10 \\
ID\_2\_898\_0    & $36\ \ \ 54.78$ & $13\ \ 50.74$ & m    & Sa   & 3 & 25.14(.20) & 0.27 & 0.49 & 0.50 &   3 \\
ID\_3\_743\_0    & $36\ \ \ 54.95$ & $12\ \ 21.44$ & E    & --   & 2 & 25.61(.20) & 0.19 & 0.11 & 0.14 &  93 \\
ID\_3\_266\_0    & $36\ \ \ 55.16$ & $13\ \ 03.60$ & E    & E    & 1 & 25.19(.19) & 0.22 & 0.30 & 0.31 &   3 \\
ID\_3\_180\_1    & $36\ \ \ 55.46$ & $13\ \ 11.19$ & E/S0 & E    & 1 & 23.94(.21) & 0.51 & 0.18 & 0.19 &  70 \\
ID\_2\_966\_0    & $36\ \ \ 55.77$ & $13\ \ 48.78$ & E    & E    & 2 & 25.83(.19) & 0.18 & 0.24 & 0.24 &  40 \\
ID\_3\_904\_0    & $36\ \ \ 55.95$ & $12\ \ 10.72$ & E    & E    & 1 & 25.11(.13) & 0.17 & 0.08 & 0.14 &  70 \\
ID\_3\_815\_1    & $36\ \ \ 56.65$ & $12\ \ 20.12$ & E/S0 & E    & 1 & 24.29(.20) & 0.41 & 0.21 & 0.22 &  11 \\
ID\_3\_355\_0    & $36\ \ \ 56.92$ & $13\ \ 01.56$ & E    & E    & 1 & 24.16(.20) & 0.39 & 0.04 & 0.08 &   5 \\
ID\_3\_726\_0    & $36\ \ \ 58.53$ & $12\ \ 33.59$ & E    & --   & 2 & 26.15(.18) & 0.18 & 0.29 & 0.28 &  12 \\
ID\_3\_363\_1    & $36\ \ \ 58.57$ & $13\ \ 05.47$ & E    & --   & 1 & 26.23(.20) & 0.11 & 0.05 & 0.04 &   5 \\
ID\_3\_813\_0    & $36\ \ \ 59.24$ & $12\ \ 27.29$ & E    & --   & 1 & 26.02(.18) & 0.17 & 0.04 & 0.08 & 135 \\
ID\_3\_748\_0    & $36\ \ \ 59.98$ & $12\ \ 35.95$ & E    & --   & 1 & 26.23(.17) & 0.15 & 0.23 & 0.20 &  16 \\
ID\_3\_888\_1    & $37\ \ \ 00.52$ & $12\ \ 25.81$ & E    & Sap  & 1 & 25.57(.18) & 0.17 & 0.12 & 0.18 &  42 \\
ID\_3\_790\_1    & $37\ \ \ 00.56$ & $12\ \ 34.60$ & E    & --   & 2 & 22.60(.14) & 0.40 & 0.10 & 0.13 &  20 \\
\hline
\hline
\end{tabular}
\end{table*}

\section{Detailed surface photometry}

The detailed surface photometry of our sample of HDF ellipticals in
the $V_{606}$ band was performed using the AIAP package (Fasano
1990), which is equipped with a completely
interactive graphical interface, allowing flexible tools for sky
subtraction, drawing, masking and fitting of the isophotes, PSF
evaluation, etc. These capabilities, as well as the allowance for
checking '{\it on--line}' all steps of the procedure, turn out to be
particularly useful when handling morphologically complex
structures and/or closely interacting objects, a situation quite
common in the case of the HDF galaxies.

Our magnitudes are given in the $STMAG$ system.

\subsection{Basic procedures}

The second release of the HDF provided us with frames '{\it ready to 
be used}', in the sense that all the standard reduction procedures
(flat fielding, bias and dark--current removal, etc.), as well as the
specific ones (drizzling, position weighting, etc..), were already 
performed and the proper (constant) sky background was subtracted 
from each frame.

Nevertheless, since not negligible and systematic residuals in the
backgrounds were detected, we decided to perform for each galaxy a
'{\it local}' refinement of the background subtraction by linear
interpolation of the residuals. After this refinement the mean
(systematic) background variations were estimated to be of the order of
$\sim 0.5\%$ of the original sky levels. These systematic variations
have been used to estimate error bars in the outer luminosity and 
geometrical profiles of each object (see next subsection).

The isophotes were drawn with a fixed surface brightness step of
$0^m.2$. Therefore, at least in the inner part of galaxies having
steep luminosity profiles, isophotes might be oversampled, since the 
difference of radius between two successive isophotes might be lower 
than the pixelsize. We note, however, that the shapes of luminosity 
profiles are not modified by the oversampling and that, whenever we
introduce luminosity profiles weighting (i.e. fitting with analytical
functions and extrapolation), we set
to zero the weight of oversampled isophotes.

Low surface brightness isophotes (usually $\mu_{606}\ge25^m.8$) were
drawn after rebinning of the frames. The AIAP package allows to
decide interactively when and how perform the rebinning, depending on
the noise of each isophote. The maximum reachable surface brightness
was not the same for all galaxies. It essentially turned out to depend 
on the presence of close companions and/or irregular sub--structures
which often make useless to go much deep in drawing the isophotes.
We can reach easily the surface brightness level of $28^m.0$. 
In the most favourable cases we were able to reach $\mu_{606}=28^m.8$. 

The isophotes of HDF ellipticals are often quite irregular due to the
intrinsic complexity of the structures (which is likely to increase at
increasing redshift), as well as to the presence of close (possibly
interacting) companions. Therefore, the flexibility of the AIAP
masking and ellipse fitting tools turned out to be useful in order to
secure reliable profiles even for very intriguing cases. We
mention, among the other things, the possibility to take fixed or
relaxed (for each isophote) the coordinates of the center of the
fitting ellipses and the possibility to complete the masked parts of the
isophotes by using the corresponding symmetric parts of the same isophotes 
with respect to the center.

\subsection{Extraction of profiles and error estimates}

Once the isophotes have been carefully masked and interpolated through
ellipses, we obtained luminosity and geometrical profiles of all
galaxies in our sample. In particular, we obtained surface brightness,
ellipticity, position angle and coefficients of the Fourier analysis
of the residuals as a function of the semi--major axes of the fitted
isophote.

The most important contribution to the uncertainties in the outer
profiles of nearby elliptical galaxies from CCD data is usually given 
by the possible errors in the estimate of the background.
In particular, an underestimation (overestimation) of the
average background level leads to a systematic distortion upwards
(downwards) of the outer luminosity profiles and then to an
overestimation (underestimation) of the total apparent luminosity. In
our case this kind of uncertainty is not the dominant one since the
average value of the background is very carefully estimated (well
outside the faint galaxy halos) by the above mentioned '{\it
local refinement}'. A more serious problem is represented by the
possible systematicity of the background residuals. Fasano and Bonoli
(1990) analyzed the influence of an artificially tilted background on
the luminosity and geometrical profiles, giving a set of formulae to
estimate error bars of surface brightness, ellipticity and position
angle of each isophote, mainly depending on the relative background
variation ($0.5\%$ in our case; see previous subsection) and on the
considered isophotal level. Error bars of our profiles were computed
according to these formulae, with an additional term (which could be
relevant for the outer isophotes) accounting for the number of points
belonging to each isophote.  This term turns out to be usually
negligible for ground--based surface photometries, where the small
signal to noise ratio of the outermost regions is largely compensated
for the great number of points defining the isophotes. In our case,
the incredible deepness and resolution of the images allowed us to
perform the detailed surface photometry of galaxies in which even the
outermost isophotes are defined by a relatively small number of
points.

\section{Results}

The profiles of surface brightness ($\mu$), ellipticity
($\varepsilon$), position angle ($\theta$) and coefficient of the
Fourier analysis of the residuals ($c_4$), as a funtion of the
semi--major axis $a$ in arcseconds ($a^{1/4}$ scale), are shown at the
end of the paper in Figure 1, together with the proper error bars (see
previus Section).  In the same figures we report for comparison the
$PSF$ profile as a dotted line. 

\subsection{Statistical properties of profiles}

\subsubsection{Luminosity profiles}

We divided our galaxy sample in 3 different classes, according to the
luminosity profiles: 

\begin{itemize}

\item most of galaxies (55) have luminosity
profiles following reasonably well the de~Vaucouleurs law up to the
innermost isophote not significantly affected by the $PSF$ (hereafter
`{\it Normal}' class);

\item a relevant fraction of galaxies (36), even following the 
de~Vaucouleurs law in the main body light distribution, exhibit in the
inner regions a substantial flattening of the luminosity profiles, not
attributable to the $PSF$ (hereafter `{\it Flat}' class);

\item finally, the isophotal analysis allowed us to detect 8 galaxies
showing complex inner structures (most of them are multi--nucleus
objects), but still obeying the de~Vaucouleurs law in the outer
profiles (hereafter `{\it Merger}' class).

\end{itemize}

In column (6) of Table 1 the luminosity profile class of each galaxy
is reported and the three classes are indicated with 1, 2 and 3,
respectively. The same convention is used in Table 2.

Concerning the `{\it Flat}' class, it is worth stressing that the
observed inward flattening of the luminosity profiles with respect to
the de~Vaucouleurs law cannot be interpreted as due to the presence of
some {\it core}--like structure. Actually, in all galaxies of this
class for which the redshift has been measured (6 objects), the linear
size of the involved regions turns out to be much greater than the
typical {\it core} size ($10^2pcs$). For instance, at $z\sim 1$,
$10^2pcs$ correspond to $\sim 0.^{\prime\prime}01$, with small
differences in the range $0.5\le z \le 3$. On the other hand, galaxies
with luminosity profile flattening confined inside $\sim
0.^{\prime\prime}1$ have been included by default in the `{\it
Normal}' class.  Moreover, by comparing the fraction of HDF
ellipticals belonging to the `{\it Flat}' class with that obtained in
a similar way from the sample of local ellipticals provided by
Djorgovski (1985), we found a relevant difference in favour of the HDF
sample ($\sim 36\%$ vs. $\sim 7\%$). We will see in a forthcoming
paper (Fasano et al. 1997) that the different classes of luminosity
profiles are often associated with different physical properties of
the galaxies (i.e. colors and sizes).

Finally, we mention that the luminosity profile of the galaxy
$ID\_2\_251\_0$ strongly suggests the presence of a nuclear point
source. This galaxy also belongs to the lists of radio and ISO sources
in the HDF (Fomalont et al. 1997, Mann et al. 1997).

\subsubsection{Ellipticity profiles}

The shape of the ellipticity profiles appears to be correlated with
the above mentioned luminosity profile classes. Actually, the galaxies
belonging to the {\it `Normal'} class show increasing or almost
constant ellipticity profiles (see Fig.1). This is the most common
behaviour in nearby early--type galaxies (see Bettoni et al. 1996).
On the other hand, 25 out of the 44 galaxies belonging to the {\it
`Flat'} class or to the {\it `Merger'} class show strongly decreasing
ellipticity profiles (see Fig.1). This is an unusual behaviour in the
local samples of early--type galaxy (Bettoni et al. 1996).  Table 2
reports some statistical data on the shapes of the ellipticity
profiles.
 
\setcounter{figure}{1}
\begin{figure}[h]
\centerline{\psfig{file=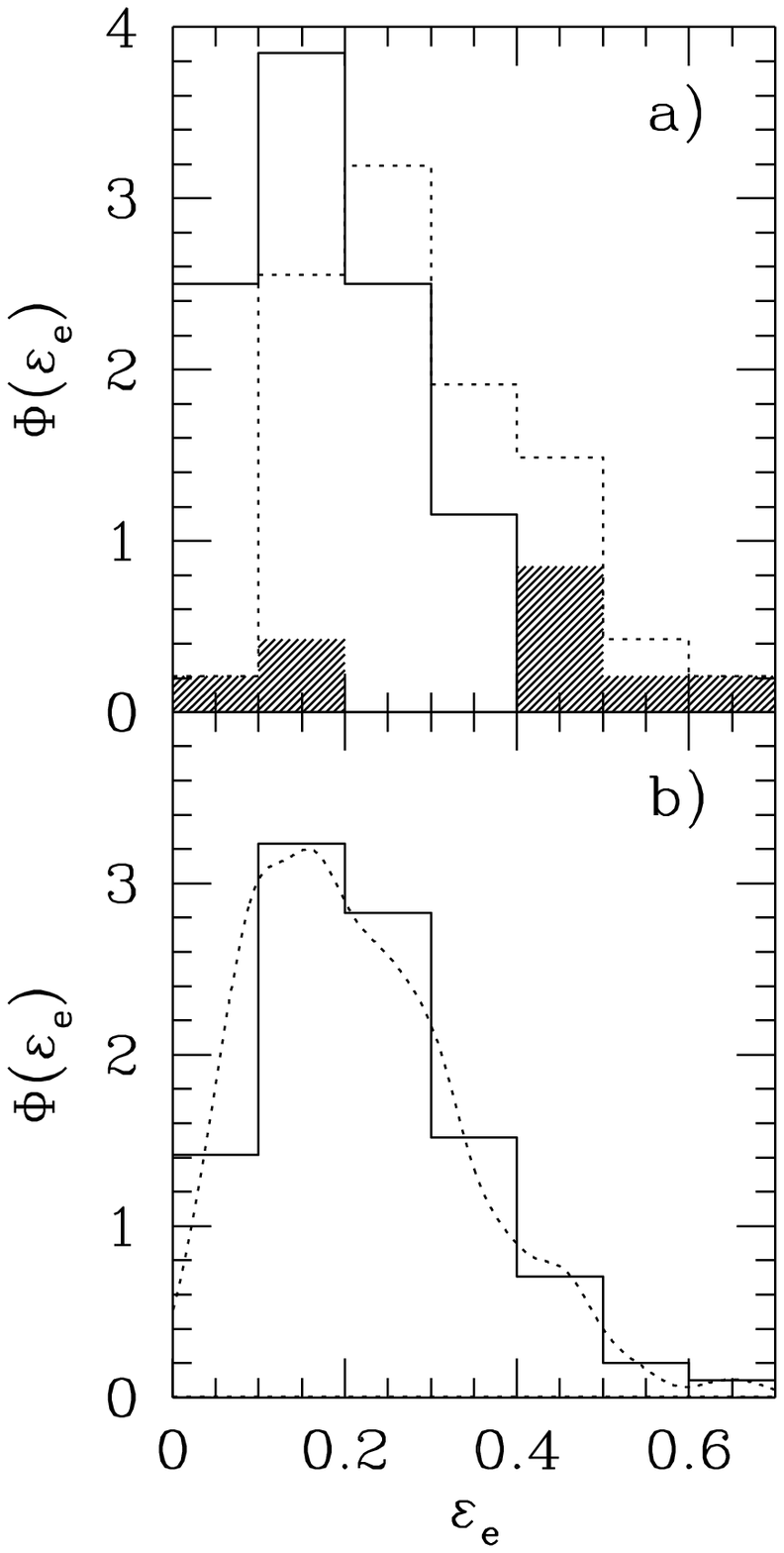,height=15.truecm,width=15.truecm}}
\caption [] {
a) effective ellipticity distribution of the 'Normal' galaxies (full 
line histogram) campared with that of 'Flat+Merger' (dotted line 
histogram). The shaded histogram refer to the 'Merger' class alone.  
b) effective ellipticity distribution of the total sample (full line 
histogram) compared with that of 'local' sample from the literature
(Fasano and Vio 1991).
}
\end{figure}

This peculiarity of the ellipticity profiles is likely to reflect on
the ellipticity distribution (computed at the effective radius, see
column 9 of Table 1) of galaxies in our sample. Figure 2a shows that
the ellipticity distribution of the {\it `Normal'} class looks
remarkably different from that relative to the other two classes, the
last ones being shifted towards flatter configurations. The
Kolmogorov--Smirnov test confirms this difference at high significance
level ($98.5\%$).  Nevertheless, the ellipticity distribution of the
whole sample (Figure 2b) seems to be in fair agreement with that
relative to nearby galaxy samples (see Fasano and Vio 1991).  It is
worth stressing that the comparison with the local samples is not
invalidated by the fact that the selection criteria of our sample
extend to the limit of recognition between stars and galaxies. In fact
the error bars shown in the ellipticity profiles take into account the
influence of the PSF (see Section 3.2).

\subsubsection{Isophotal shape and twisting}

In Table 2 we report some statistics on the shape of the isophotes
(`{\it disky}' for $c_4 > 0$; `{\it boxy}' for $c_4 < 0$) in our
galaxy sample for the different luminosity profile classes.  There is
a weak indication than the fraction of {\it boxy} galaxies increases
from the `{\it Normal}' to the `{\it Flat}' and `{\it Merger}'
classes.

\setcounter{figure}{2}
\begin{figure}[t]
\centerline{\psfig{file=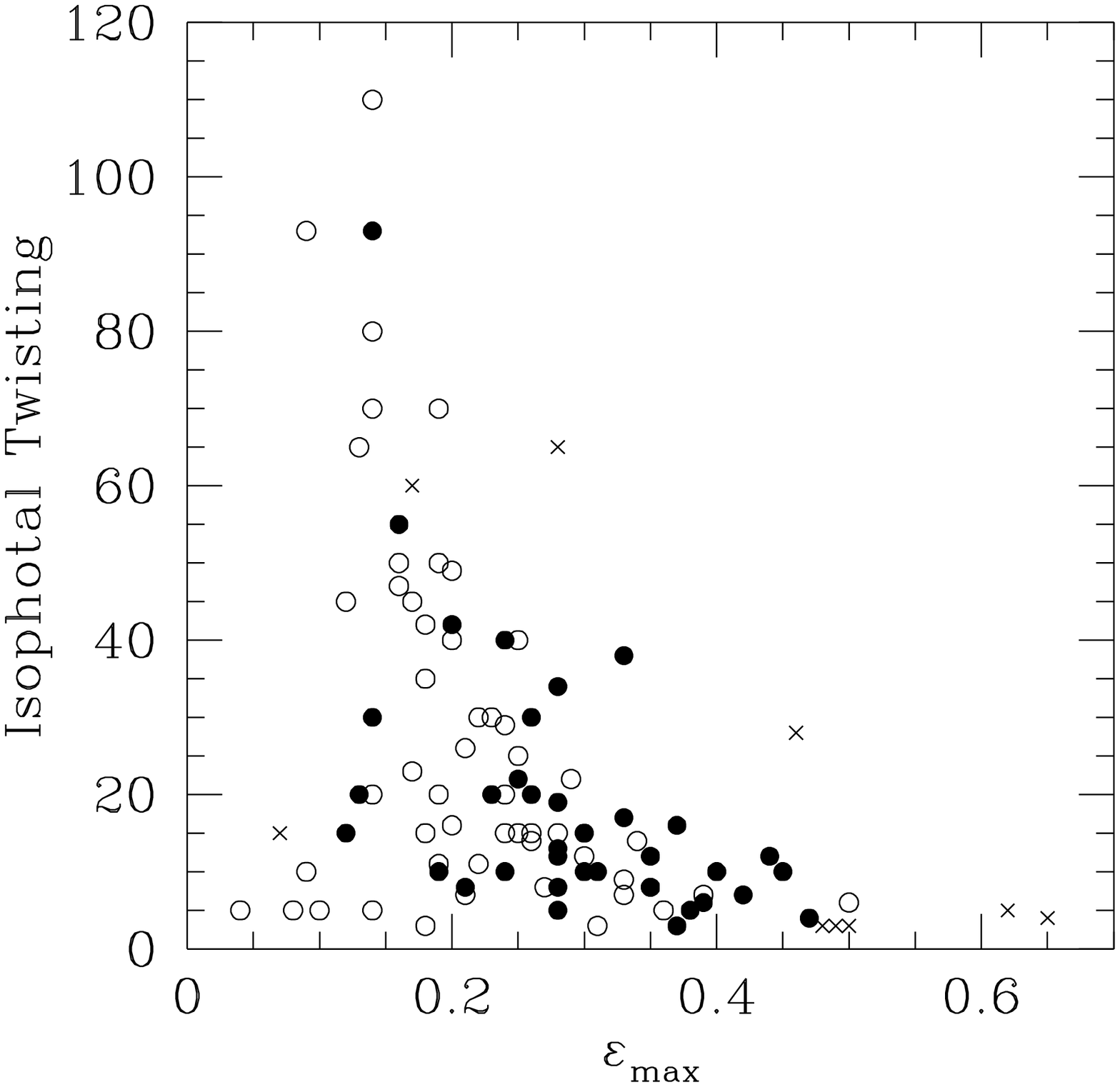,height=9.truecm,width=8.truecm}}
\caption [] {
Maximum ellipticity vs. isophotal twisting (open circles: 'Normal' 
class; full circles: 'Flat' class; crosses: 'Merger' class).
}
\end{figure}

Concerning the position angle profiles, although the uncertainties
involved in measuring the position angle of outer isophotes are
relevant for HDF ellipticals, we estimate that the amount of isophotal
twisting in our sample is larger (on average) than that found in local
galaxy samples (see Fasano and Bonoli 1989). This fact may be
explained as a consequence of the high fraction of morphologically
perturbed objects in the HDF.

Figure 3 shows the distribution of $HDF$ early--type
galaxies in the `{\it maximum ellipticity - twisting}' plane, which,
even being qualitatively similar to that of local samples (see
Galletta 1980), shows an higher fraction of significantly twisted
objects. The maximum ellipticity $\varepsilon_{max}$ and the total
isophotal twisting (in degrees) are reported in columns (10) and (11)
of Table 1, respectively.
 
\setcounter{figure}{3}
\begin{figure*}[t]
\centerline{\psfig{file=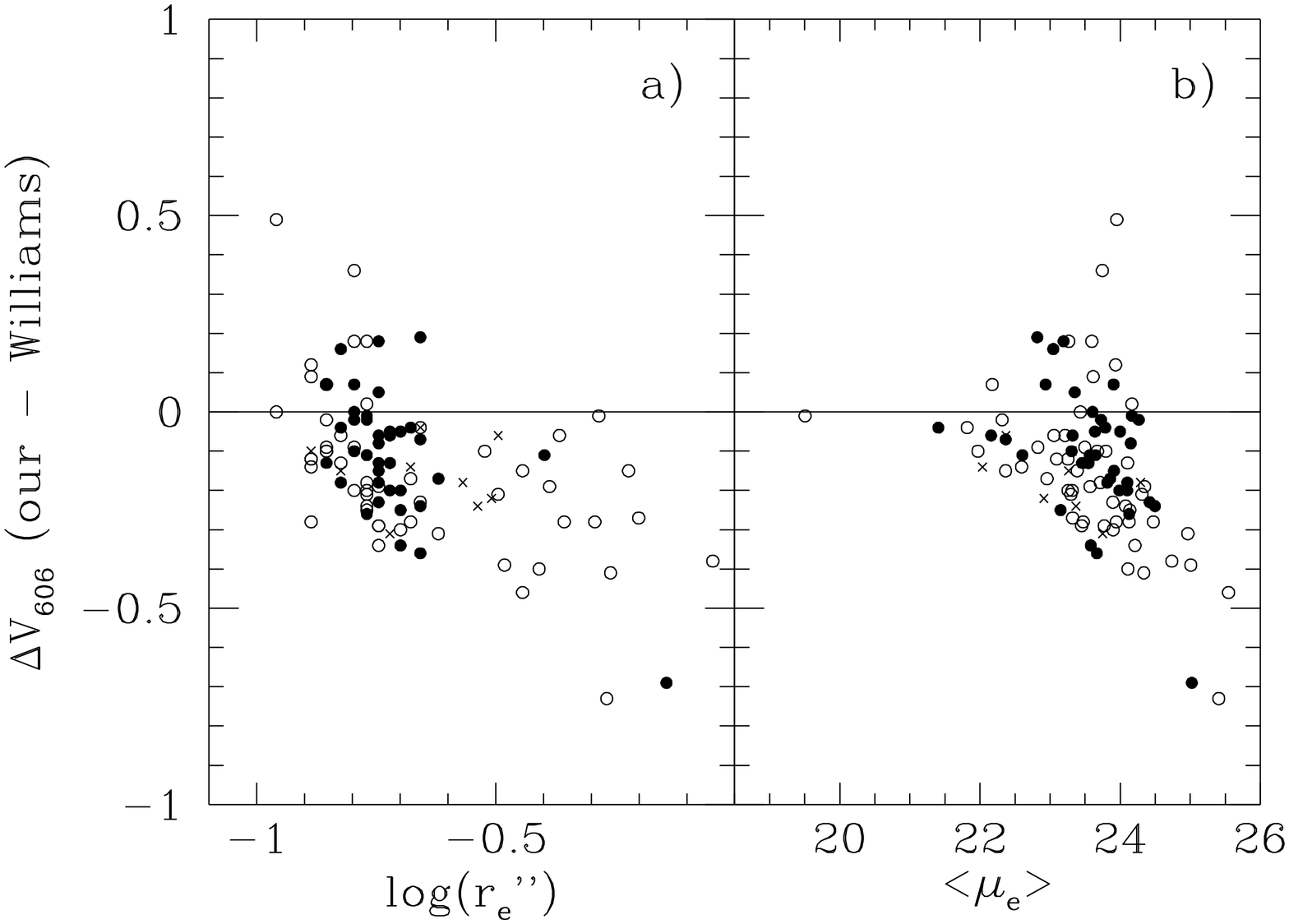,height=7.0truecm,width=12.truecm,angle=-90}}
\caption [] {
Difference between our total magnitudes and the magnitudes given
by Williams et al. (1996) as a function of the effective radius
(panel a) and of the average surface brightness (panel b).
The meaning of the symbols is the same as in Fig.3.
}
\end{figure*}

\begin{table}
\caption[ ]{Statistical Properties of profiles}
\begin{tabular}{lccc|c}
\hline
\hline
Class & 1 & 2 & 3 & All \\
\hline
\multicolumn{5}{c}{average ellipticities} \\
\hline 
$<\varepsilon_e>$ & 0.17 & 0.28 & 0.38 & 0.23 \\
$<\varepsilon_{max}>$ & 0.21 & 0.30 & 0.40 & 0.26 \\
\hline
\multicolumn{5}{c}{ellipticity profiles ($\%$)} \\
\hline
$\varepsilon\nearrow$ & 36 & 14 & 0 & 25 \\
$\varepsilon\sim$ & 58 & 33 & 22 & 46 \\
$\varepsilon\searrow$ & 6 & 53 & 75 & 29 \\
\hline
\multicolumn{5}{c}{isophotal shape ($\%$)} \\
\hline 
{\it disky} & 33 & 25 & 11 & 28 \\
{\it elliptical} & 14 & 11 & 0 & 12 \\
{\it boxy} & 22 & 33 & 50 & 29 \\
{\it irregular} & 31 & 31 & 33 & 31 \\
\hline
\#gal. & 55 & 36 & 8 & 99 \\
\hline
\hline
\end{tabular}
\end{table}

\subsection{Extraction of the global parameters}

In order to derive total magnitudes and half--light radii of the
galaxies, it is convenient to use some analytical representation
of the luminosity profiles. This allows a suitable smoothing
of each profile and provides an easy way to extrapolate it.
For obvious reasons it is also convenient to operate on the 
`{\it equivalent}' luminosity profiles, that is to multiply the
semi--major axis ($a$) by the factor $\sqrt{1-\varepsilon (a)}$.

The most common technique to get suitably smooth representations of
any observed function is the bicubic--spline interpolation. In the
case of luminosity profiles of elliptical galaxies, a nice
representation is also given by the Sersic function (1968, see also
Ciotti 1991). However, we preferred to represent the equivalent
luminosity profiles by means of sums of gaussian functions whose peak
intensities regularly decrease at increasing the standard deviations
(multi--gaussian expansion technique). In a forthcoming paper (Fasano
et al. 1997) we will see that this representation is useful to perform
the deconvolution of the luminosity profiles (Bendinelli 1991,
Emsellem et al. 1994). Here we wish only to mention that, in our
particular case, this method gives usually a better representation
with respect to the bicubic--spline method, especially in the inner
part of luminosity profiles.

The multi--gaussian representation was also used to extrapolate the
profiles. In general, it was forced to follow the de~Vaucouleurs law
down to very faint values of $\mu$. However, if the galaxy size is
comparable with the $PSF$ size (very steep profiles), it is necessary
to impose that the multi--gaussian extrapolation of the outer galaxy
profile does not fall below the very extended wings of the  
$HST$-$PSF$ itself. To this end we
forced the extrapolation of the luminosity profiles of very small
galaxies to converge smoothly towards the $PSF$ profile at large
radii.  Luminosity profiles with an effective radius larger than three
times the $FWHM$ were extrapolated simply by a de~Vaucouleurs' law.

A special warning is needed when computing the total magnitude of
galaxies belonging to the above defined `{\it Merger}' class, since
their complex inner structures make undefined (or unreliable) the
inner part of luminosity profiles. In these cases the flux inside the
innermost reliable ellipse was directly mesured on the frame and was
considered as an additional contribution to the integral of the
luminosity profile, computed from that ellipse
and extrapolated by a de~Vaucouleurs law.
The total magnitudes in the $V_{606}$ band ($STMAG$ system) and the
corresponding equivalent effective (half--light) radii $r_e$ are
reported in the columns (7) and (8) of Table 1, respectively.
The quantities in brackets close to the column (7) represent 
the surplus magnitudes $\Delta_m$ due to the extrapolation procedure. 
They give an indication of the quality of the total magnitude estimates.
Adding these quantities to $V^{Tot}_{606}$ one obtain, for
each galaxy, the magnitude before extrapolation.

In Figure 4 we compare the $V_{606}$-$STMAG$ total magnitudes from our
detailed surface photometry with the automated $FOCAS$ magnitudes
given by Williams et al. (1996), corrected for the average offset of
$0^m.2$ between the $ABMAG$ and the $STMAG$ systems. There are two
galaxies ($ID\_2\_726\_1$ and $ID\_4\_289\_0$) for which the flux has
been probably overestimated by $FOCAS$, due to the presence of very
close companions. Apart from these cases, the total fluxes computed
with our procedure tend to be systematically greater than those
obtained with the automated photometry. Moreover, the difference
increases at increasing both the average surface brightness of the
galaxies and (weakly) their angular size. This fact is not surprising
and it is likely to indicate that the automated photometry tends to
underestimate the halos of the ellipticals. To this concern, it is also 
worth noticing that the differences $\Delta V_{mag}$ in Figure 4 turn out 
to be roughly proportional to the previously mentioned quantities 
$\Delta_m$.

\clearpage
\setcounter{figure}{0}
\begin{figure*}
\centerline{\psfig{file=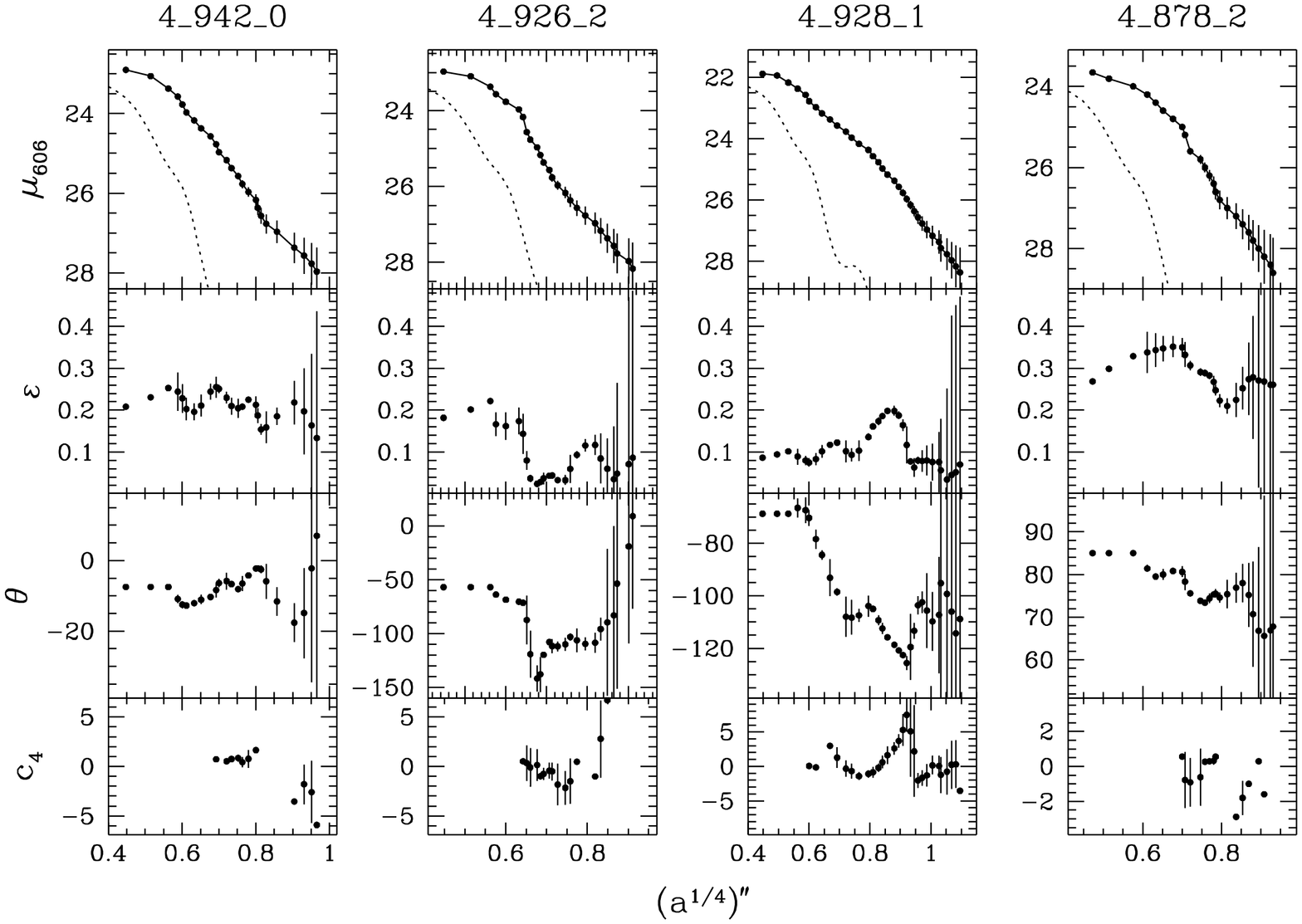,height=11.truecm,width=19.truecm,angle=-90}}
\end{figure*}
\begin{figure*}
\centerline{\psfig{file=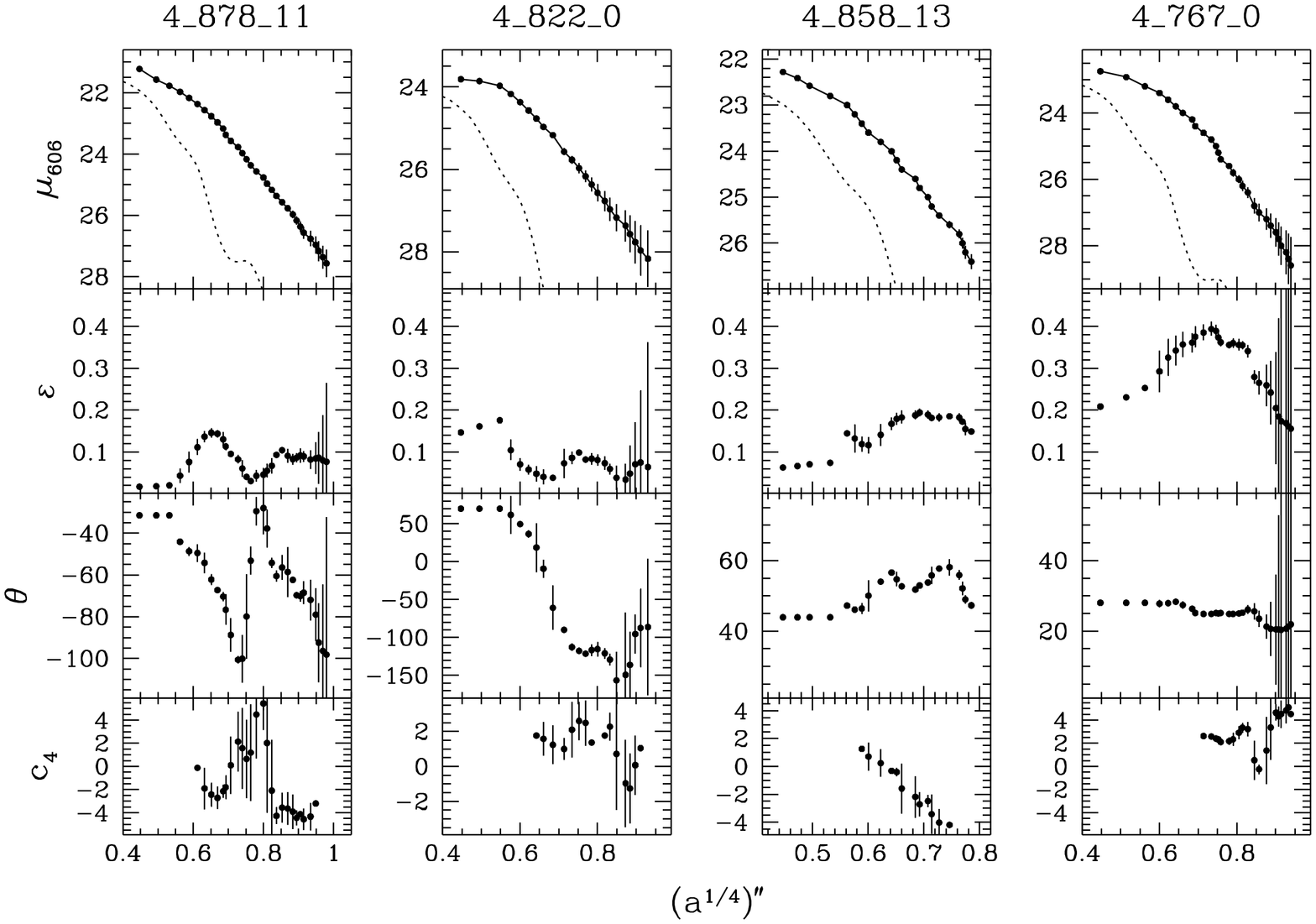,height=11.truecm,width=19.truecm,angle=-90}}
\caption [] {
Luminosity and geometrical profiles of elliptical galaxies in the HDF}
\end{figure*}

\setcounter{figure}{0}
\begin{figure*}
\centerline{\psfig{file=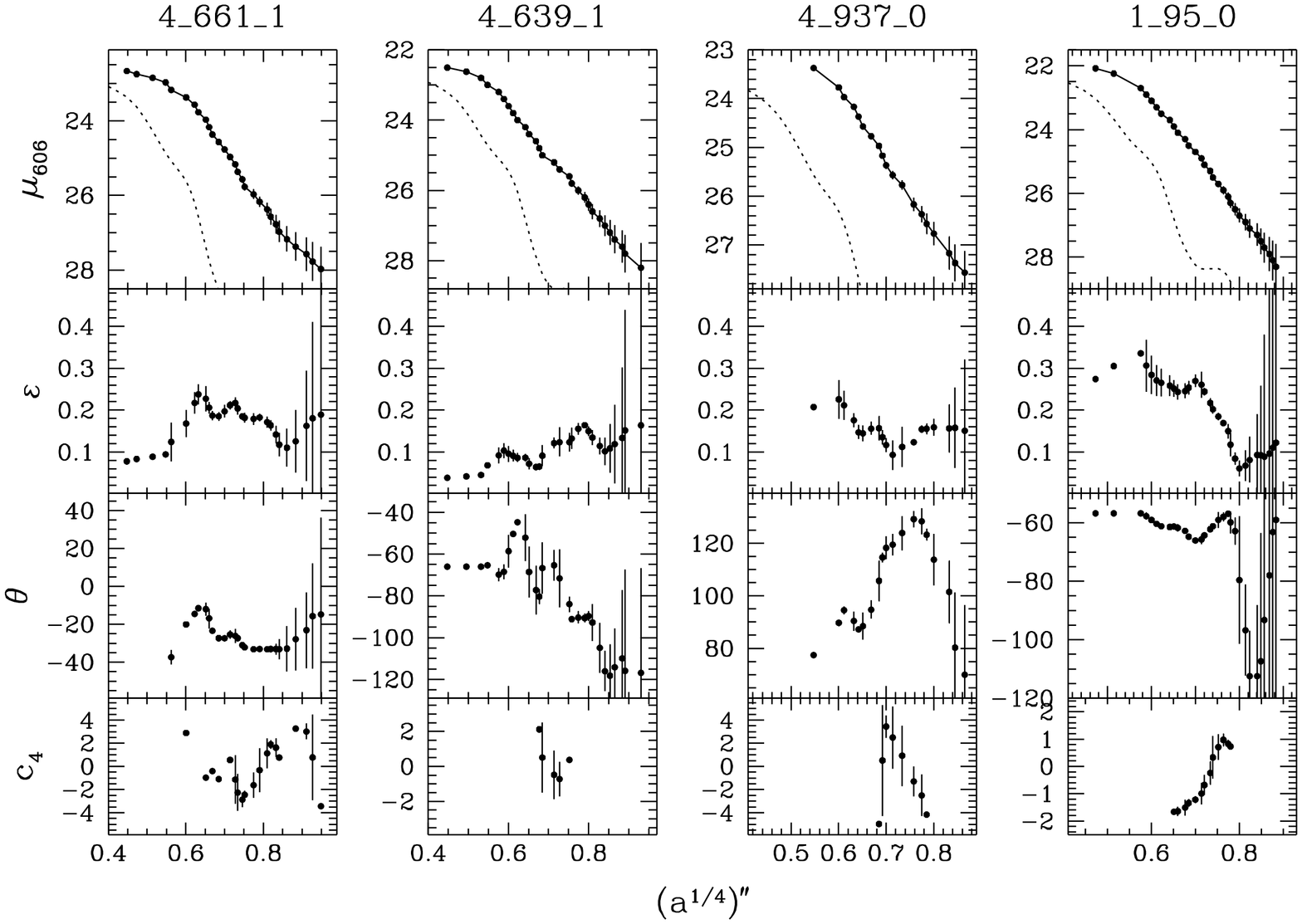,height=11.truecm,width=19.truecm,angle=-90}}
\end{figure*}
\begin{figure*}
\centerline{\psfig{file=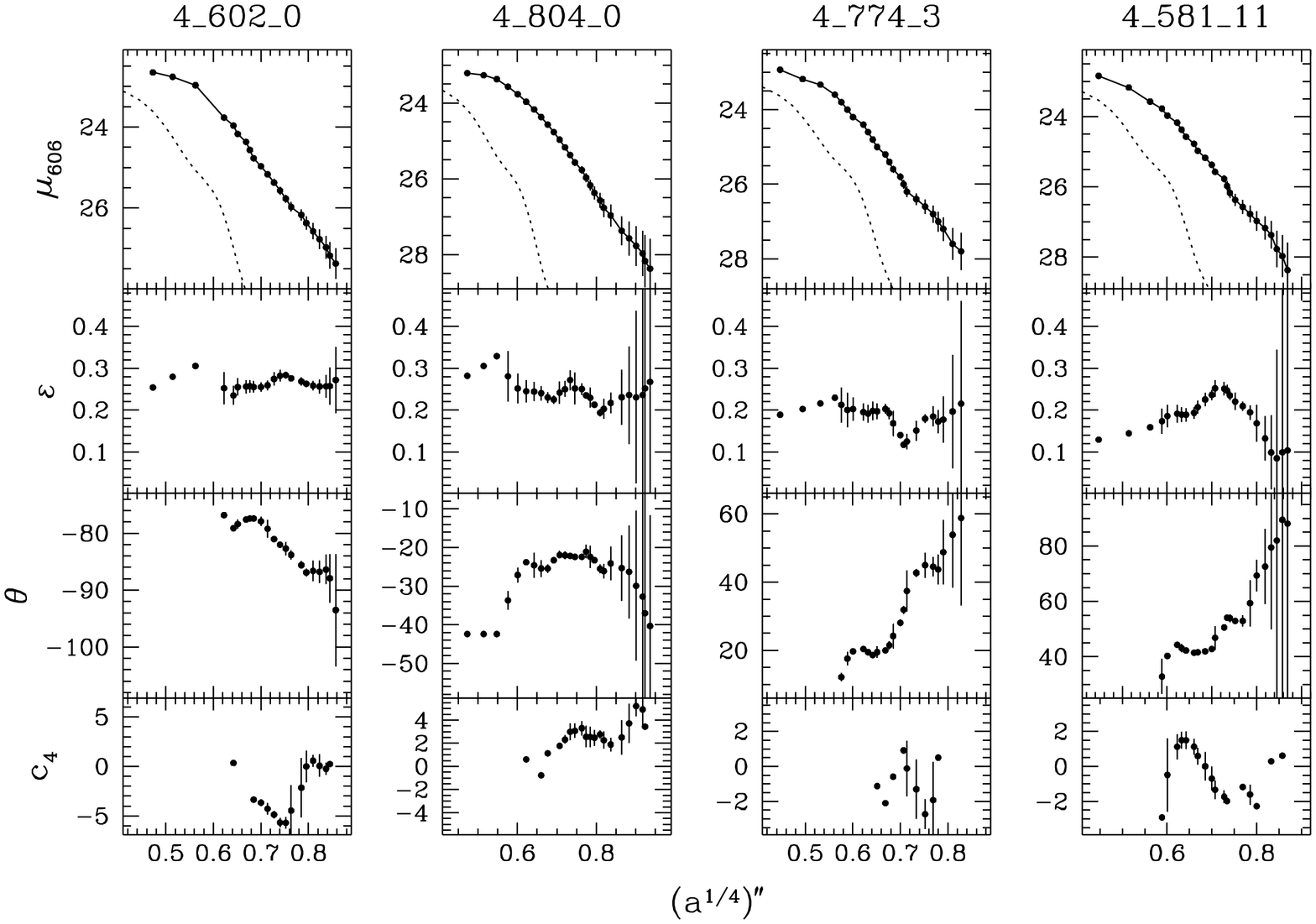,height=11.truecm,width=19.truecm,angle=-90}}
\caption [] {
...continue...}
\end{figure*}

\setcounter{figure}{0}
\begin{figure*}
\centerline{\psfig{file=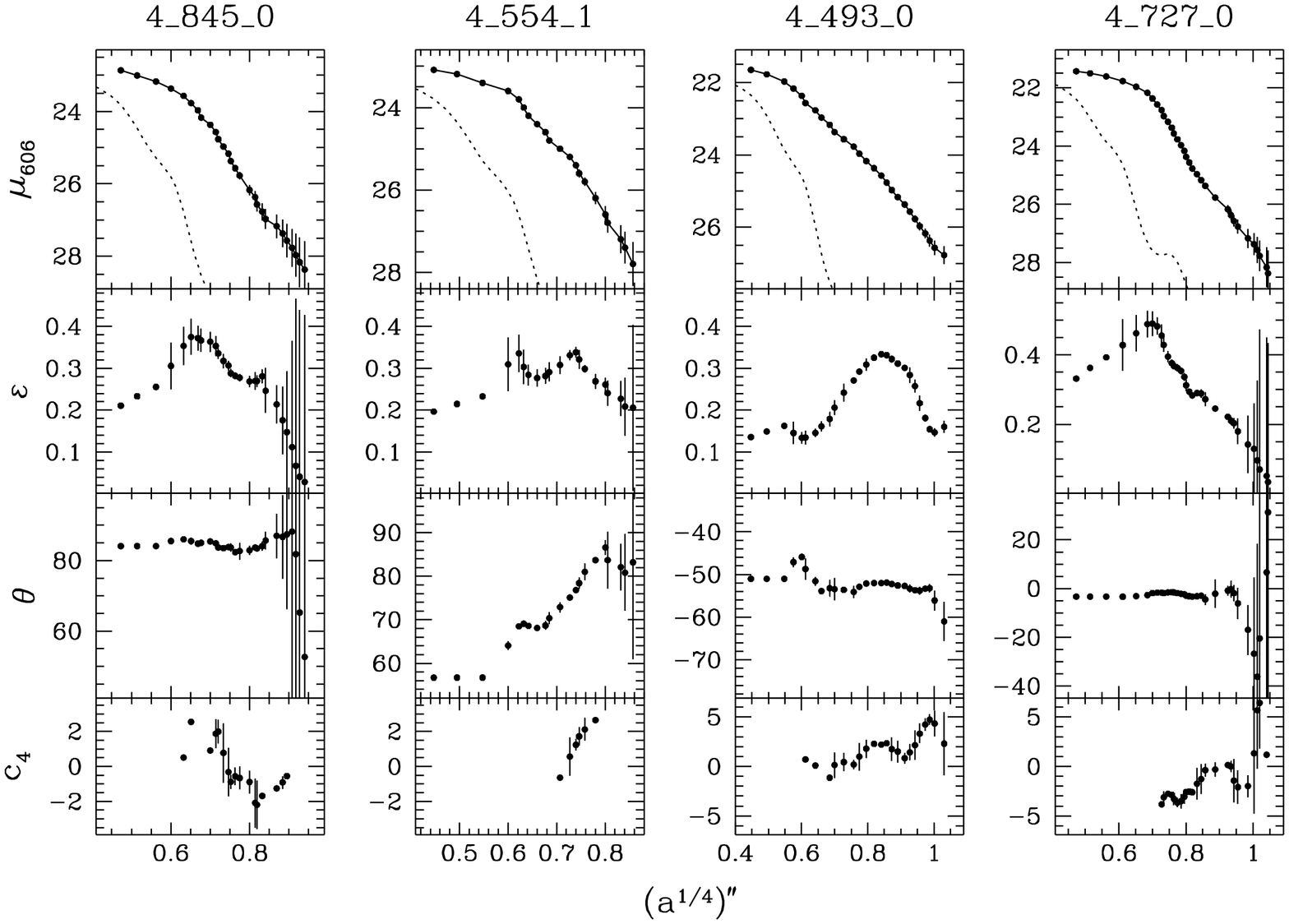,height=11.truecm,width=19.truecm,angle=-90}}
\end{figure*}
\begin{figure*}
\centerline{\psfig{file=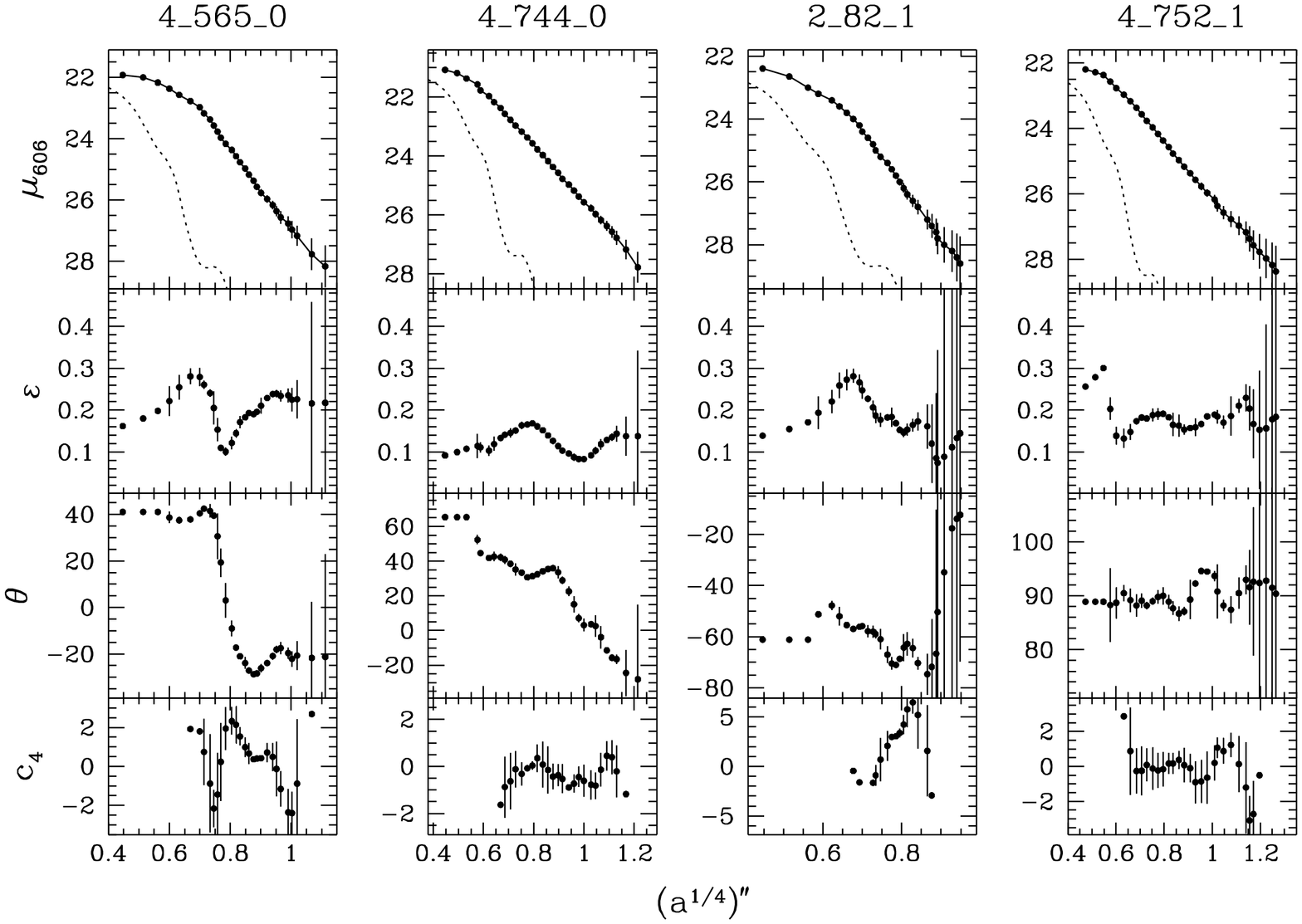,height=11.truecm,width=19.truecm,angle=-90}}
\caption [] {
...continue...}
\end{figure*}

\setcounter{figure}{0}
\begin{figure*}
\centerline{\psfig{file=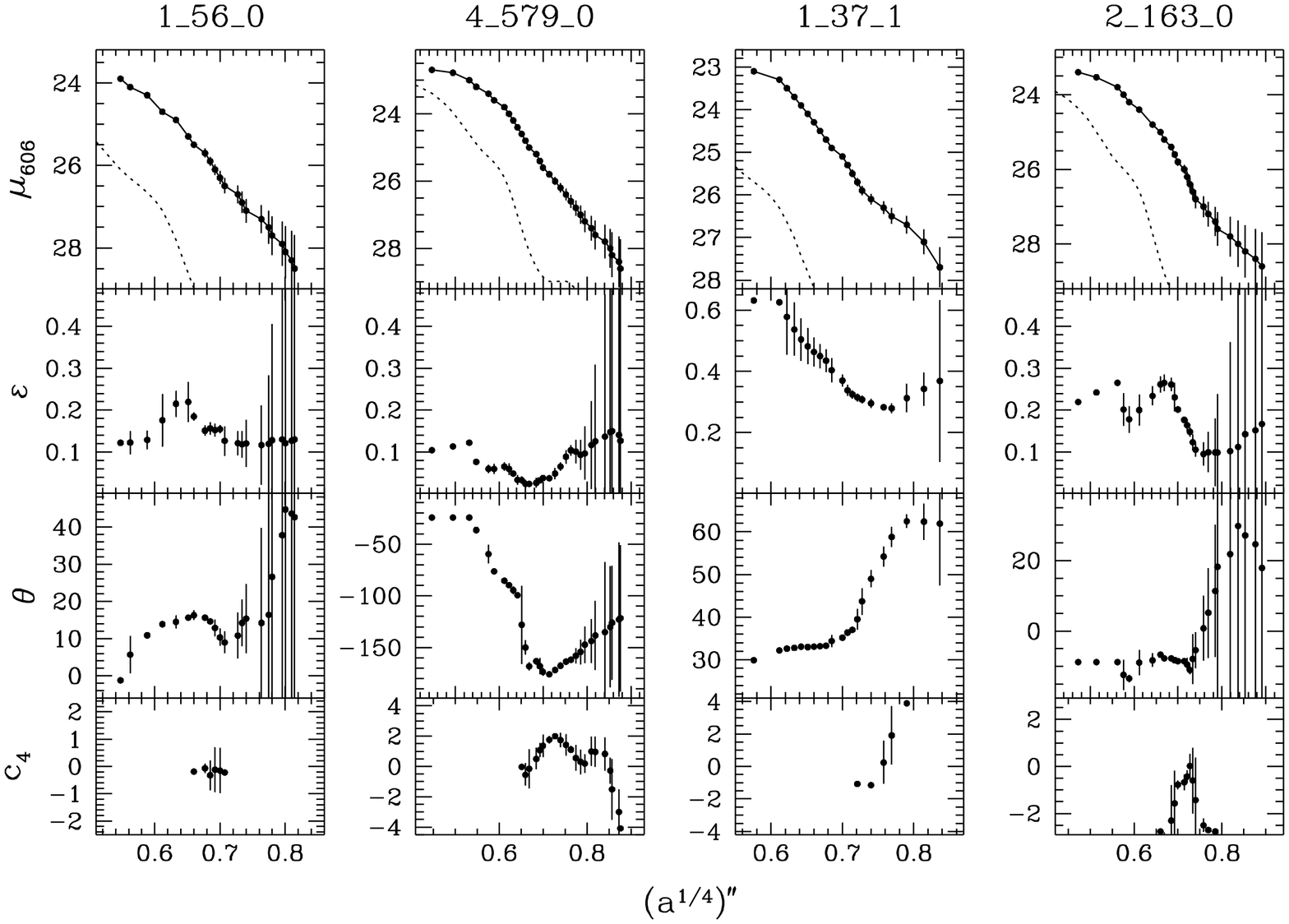,height=11.truecm,width=19.truecm,angle=-90}}
\end{figure*}
\begin{figure*}
\centerline{\psfig{file=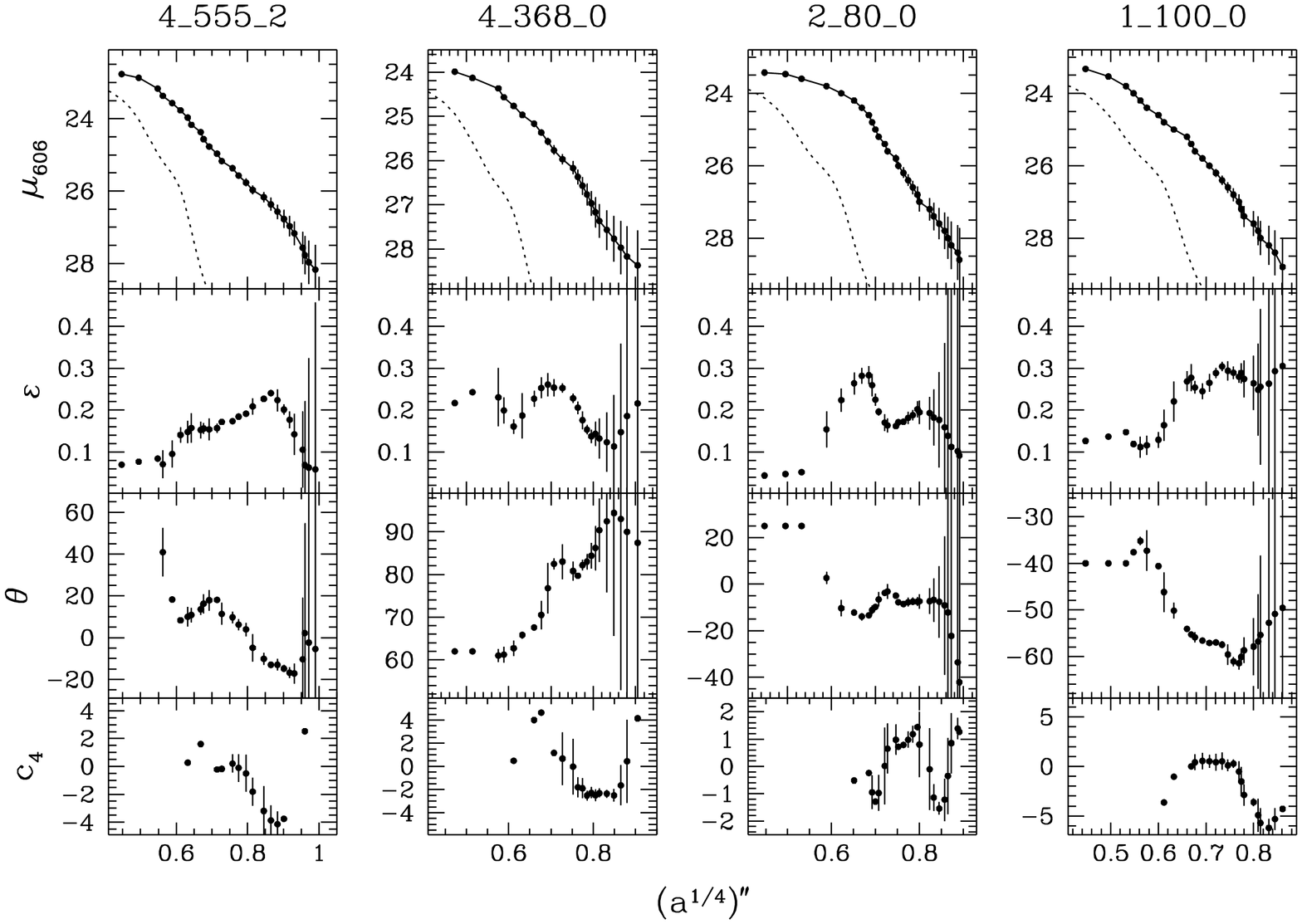,height=11.truecm,width=19.truecm,angle=-90}}
\caption [] {
...continue...}
\end{figure*}

\setcounter{figure}{0}
\begin{figure*}
\centerline{\psfig{file=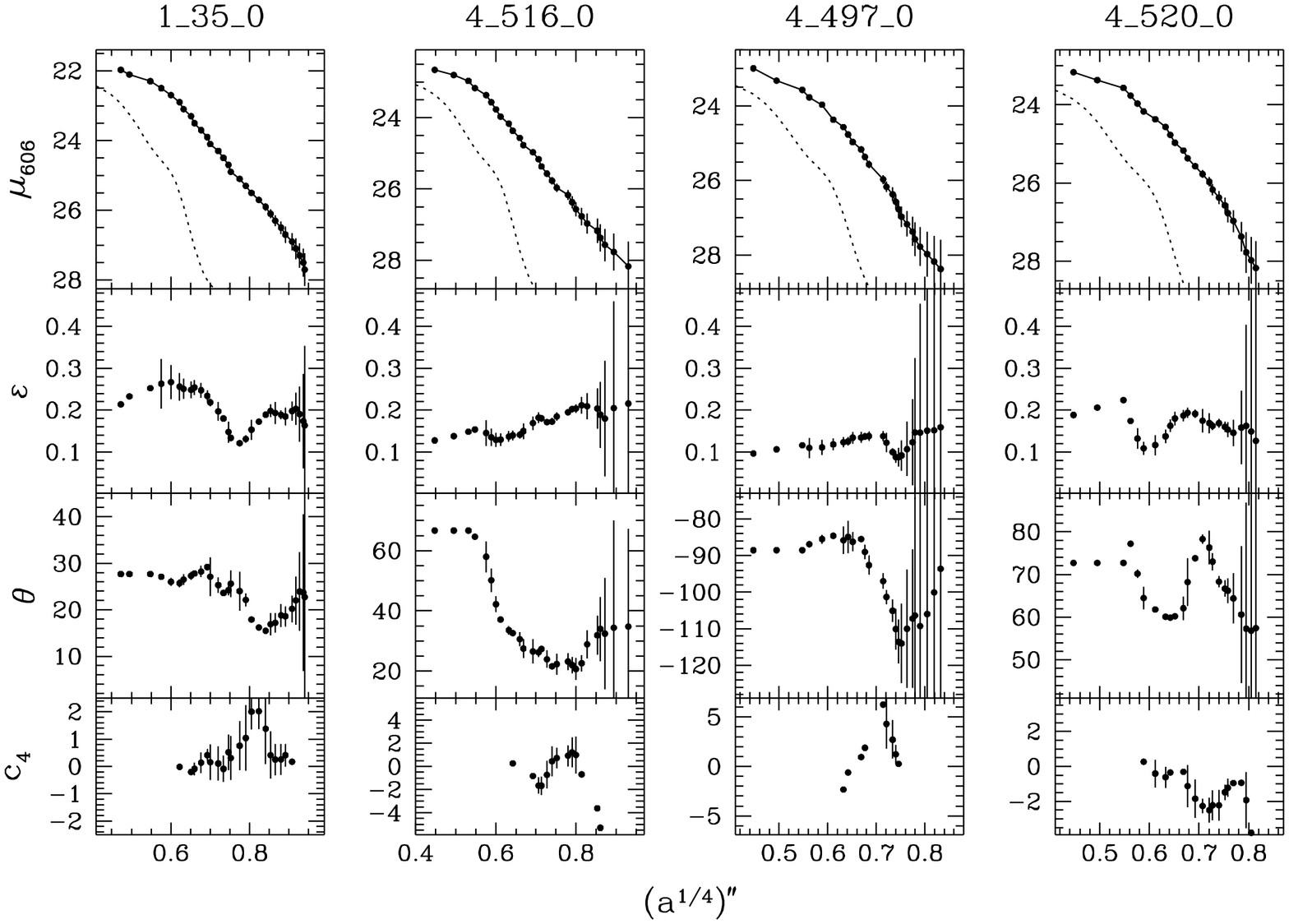,height=11.truecm,width=19.truecm,angle=-90}}
\end{figure*}
\begin{figure*}
\centerline{\psfig{file=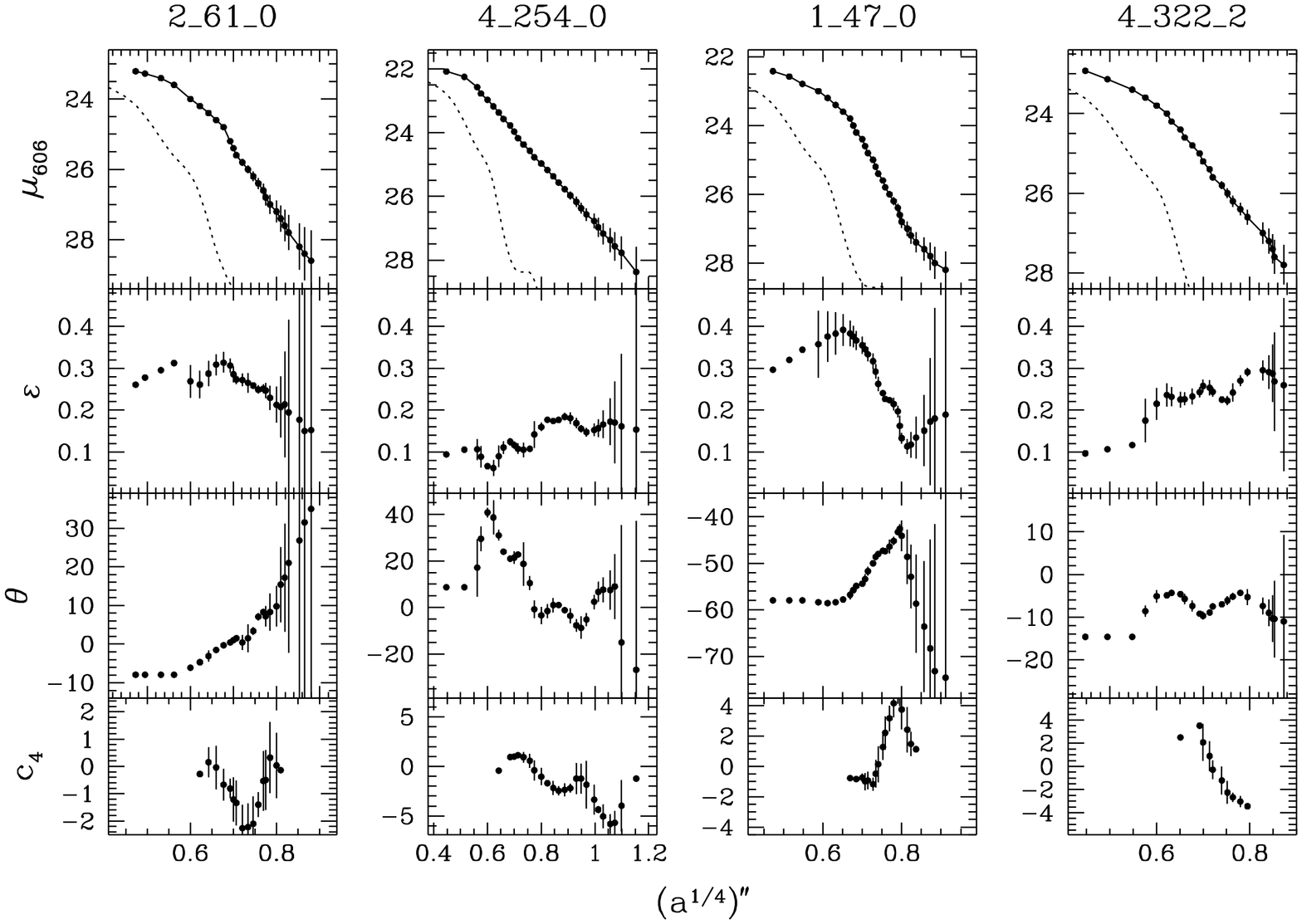,height=11.truecm,width=19.truecm,angle=-90}}
\caption [] {
...continue...}
\end{figure*}

\setcounter{figure}{0}
\begin{figure*}
\centerline{\psfig{file=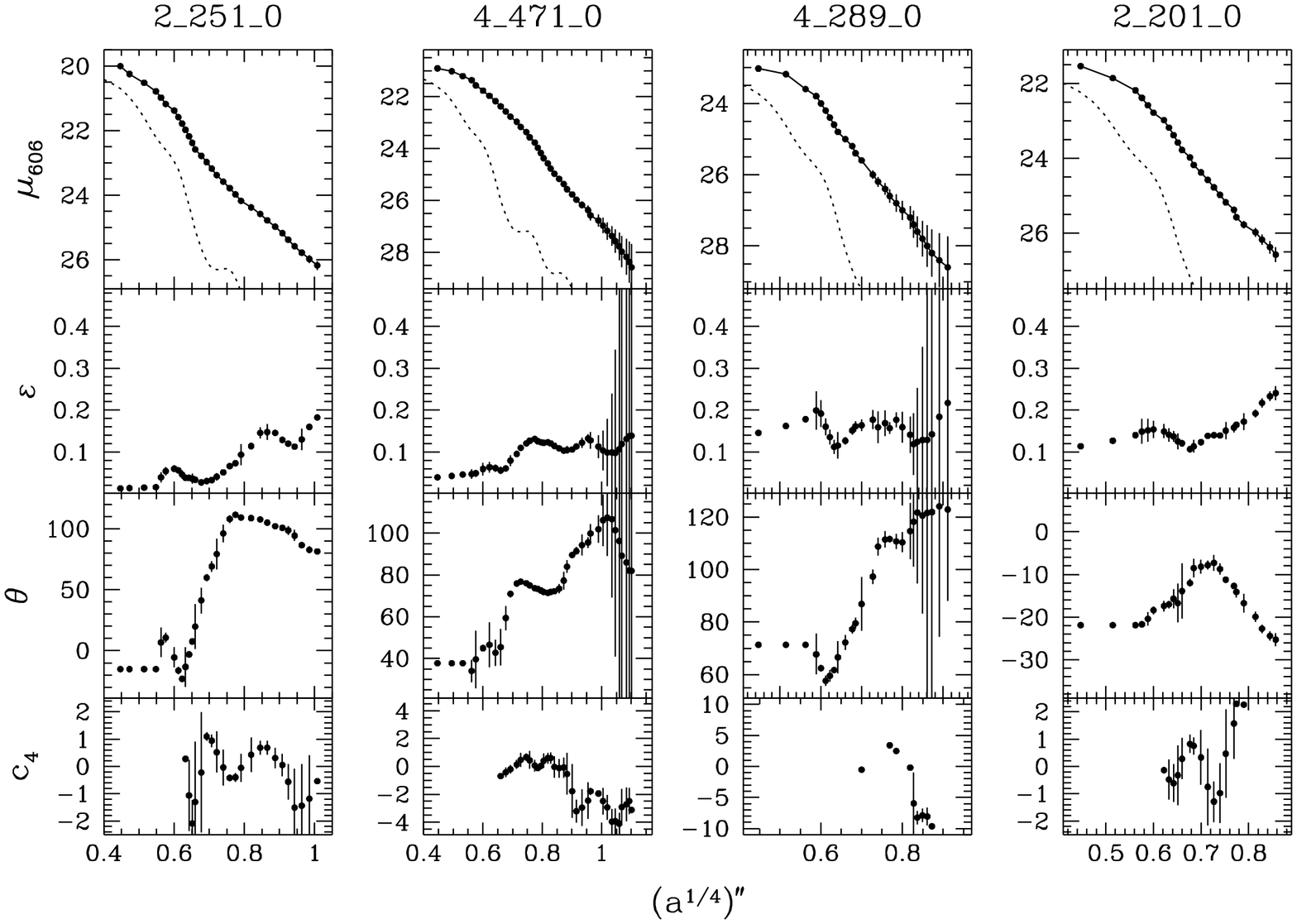,height=11.truecm,width=19.truecm,angle=-90}}
\end{figure*}
\begin{figure*}
\centerline{\psfig{file=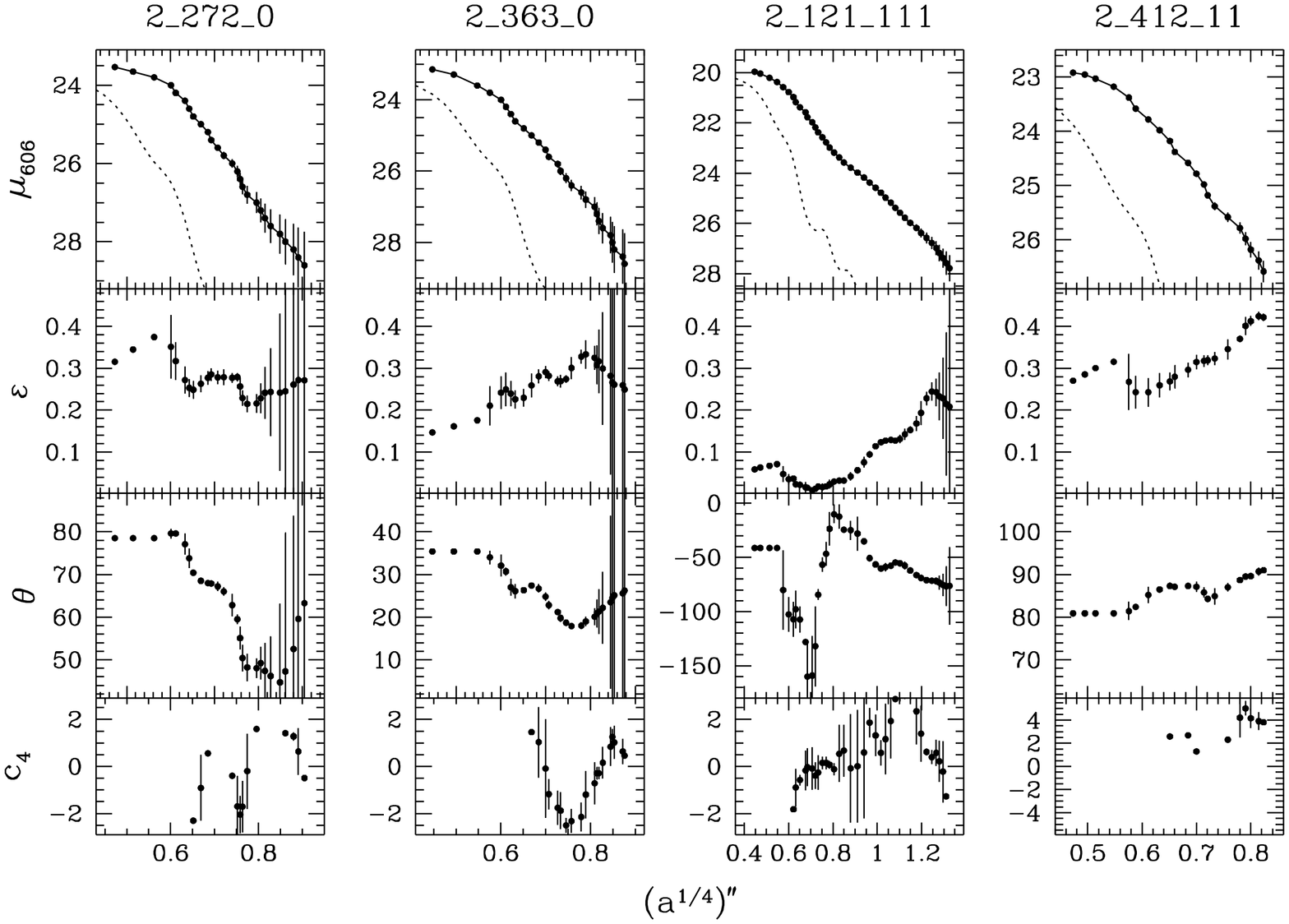,height=11.truecm,width=19.truecm,angle=-90}}
\caption [] {
...continue...}
\end{figure*}

\setcounter{figure}{0}
\begin{figure*}
\centerline{\psfig{file=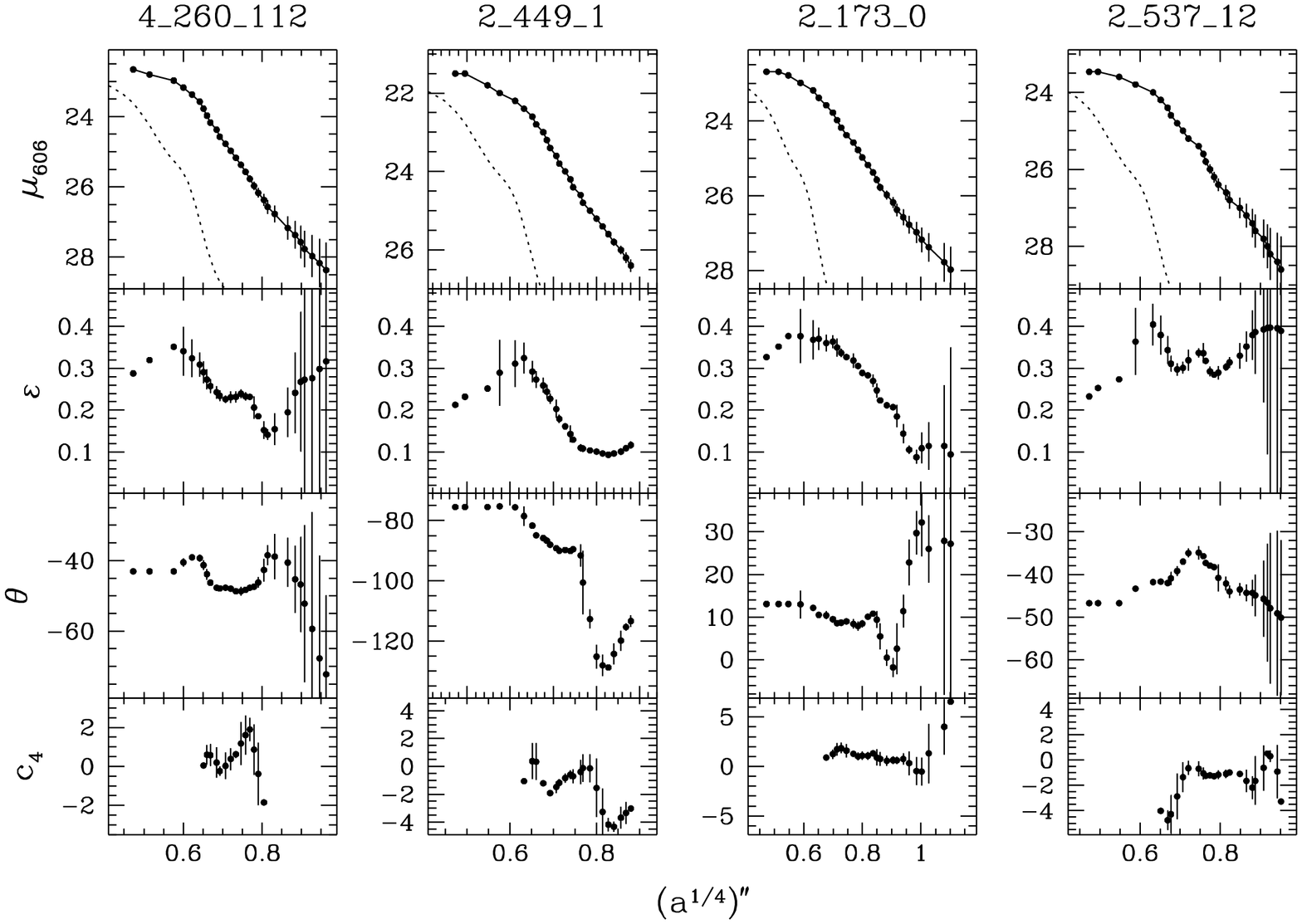,height=11.truecm,width=19.truecm,angle=-90}}
\end{figure*}
\begin{figure*}
\centerline{\psfig{file=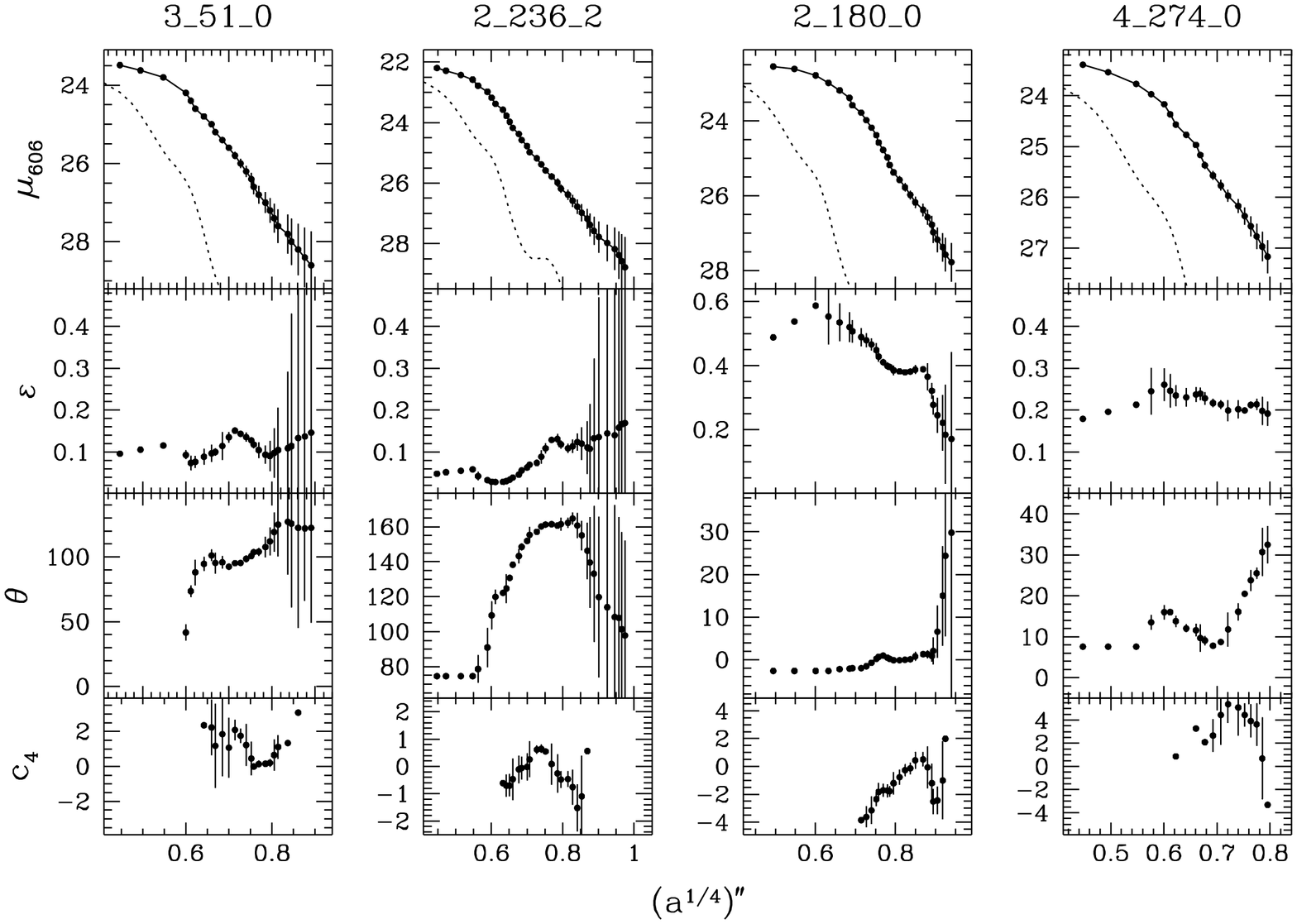,height=11.truecm,width=19.truecm,angle=-90}}
\caption [] {
...continue...}
\end{figure*}

\setcounter{figure}{0}
\begin{figure*}
\centerline{\psfig{file=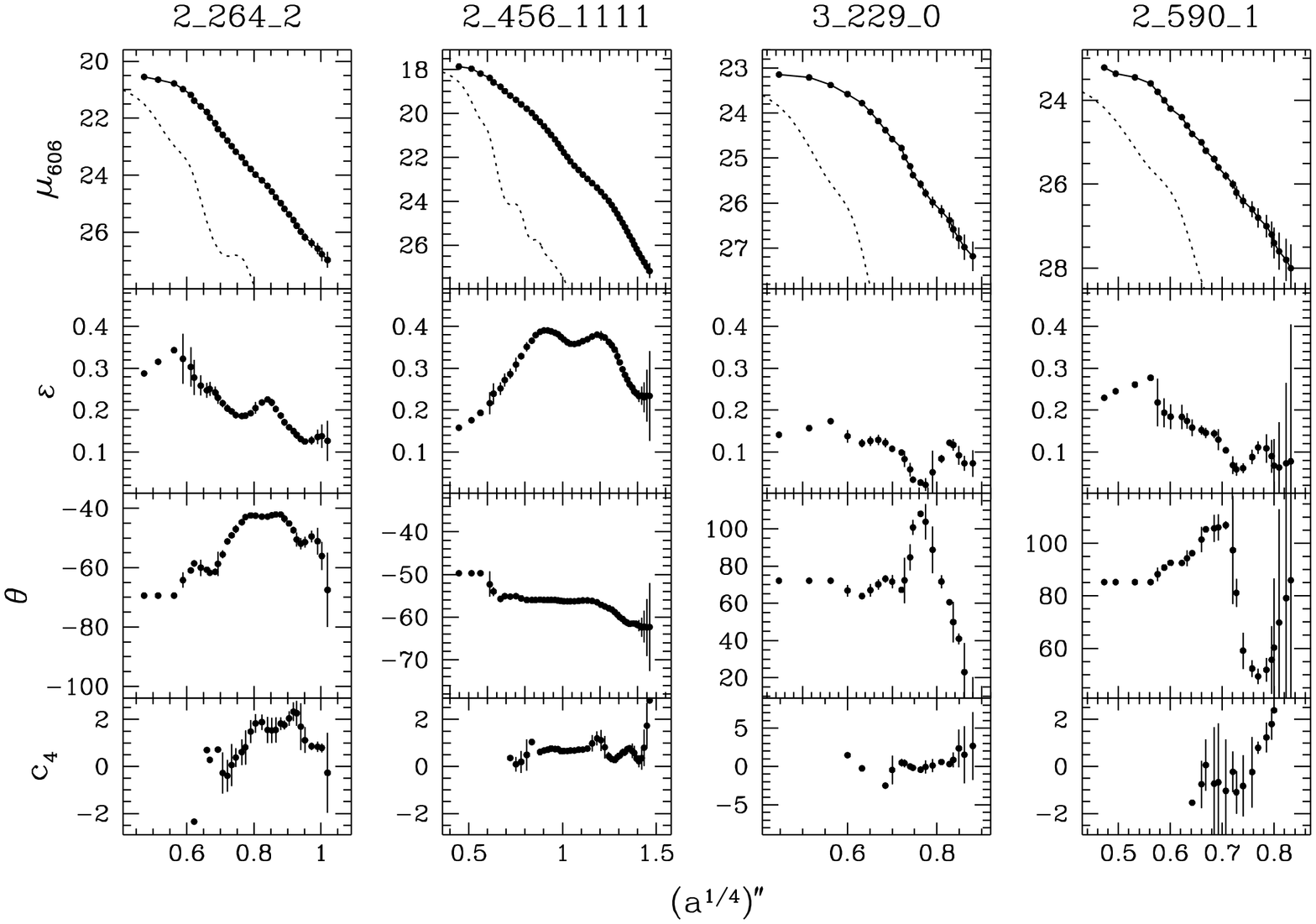,height=11.truecm,width=19.truecm,angle=-90}}
\end{figure*}
\begin{figure*}
\centerline{\psfig{file=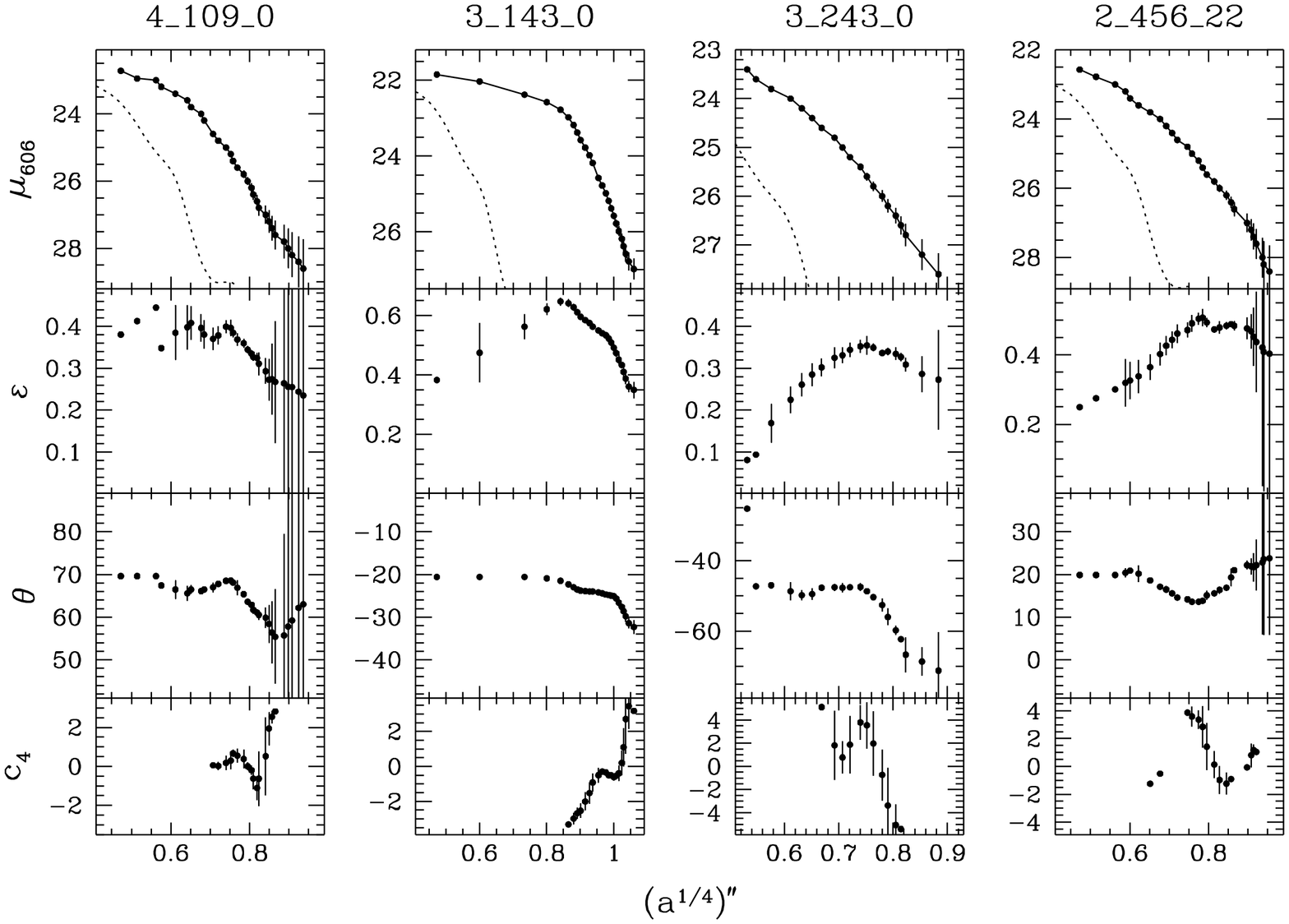,height=11.truecm,width=19.truecm,angle=-90}}
\caption [] {
...continue...}
\end{figure*}

\setcounter{figure}{0}
\begin{figure*}
\centerline{\psfig{file=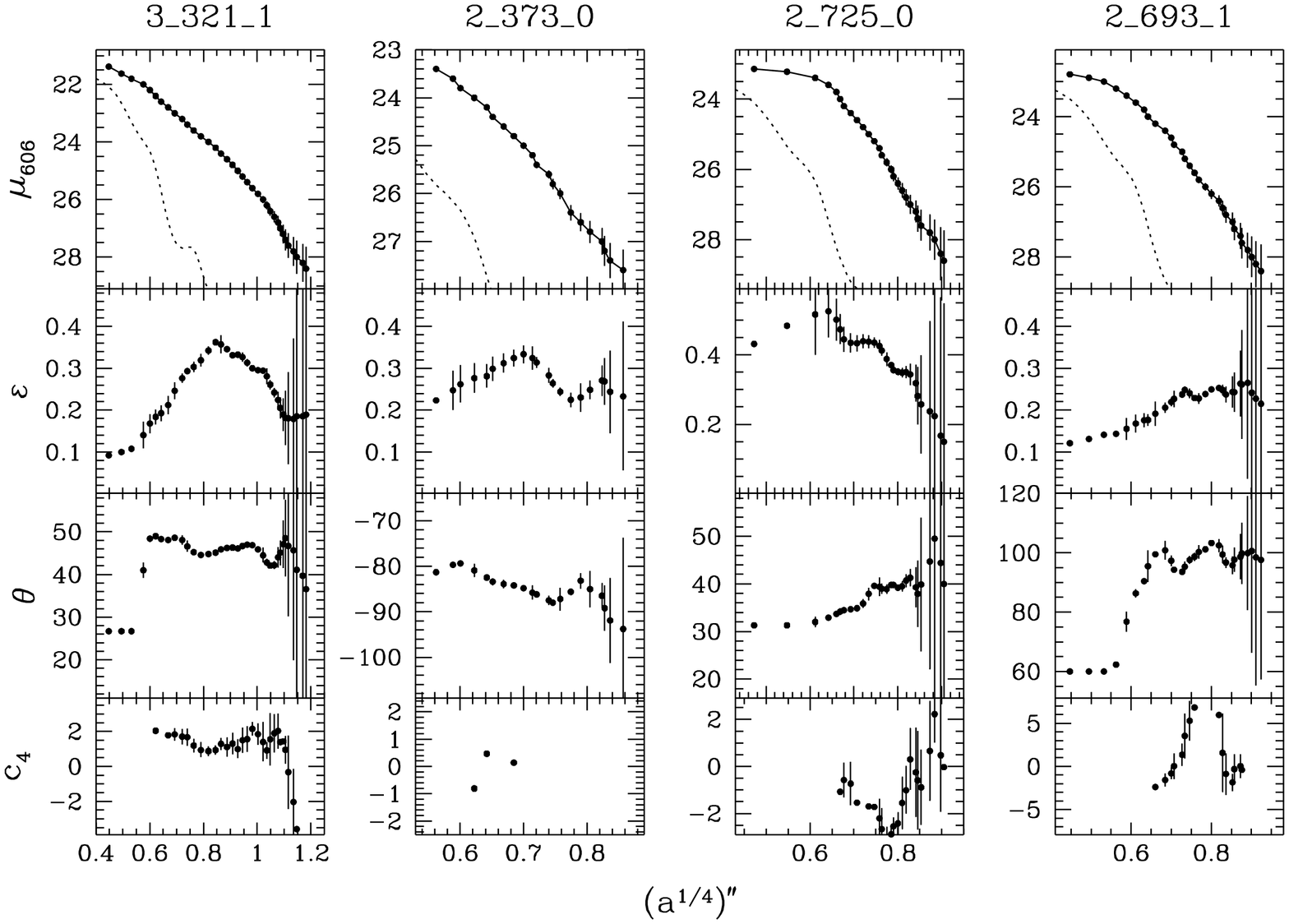,height=11.truecm,width=19.truecm,angle=-90}}
\end{figure*}
\begin{figure*}
\centerline{\psfig{file=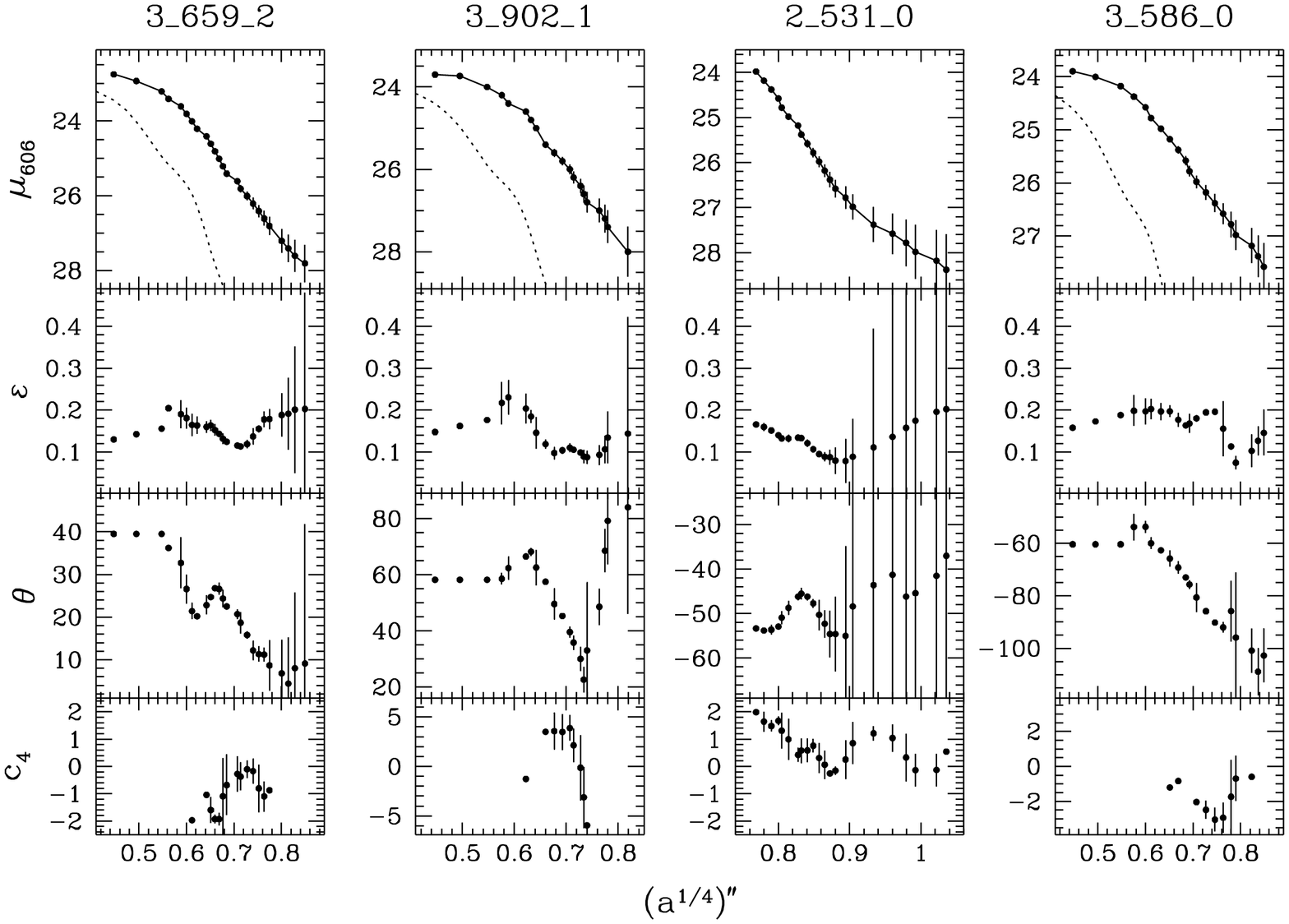,height=11.truecm,width=19.truecm,angle=-90}}
\caption [] {
...continue...}
\end{figure*}

\clearpage\setcounter{figure}{0}
\begin{figure*}
\centerline{\psfig{file=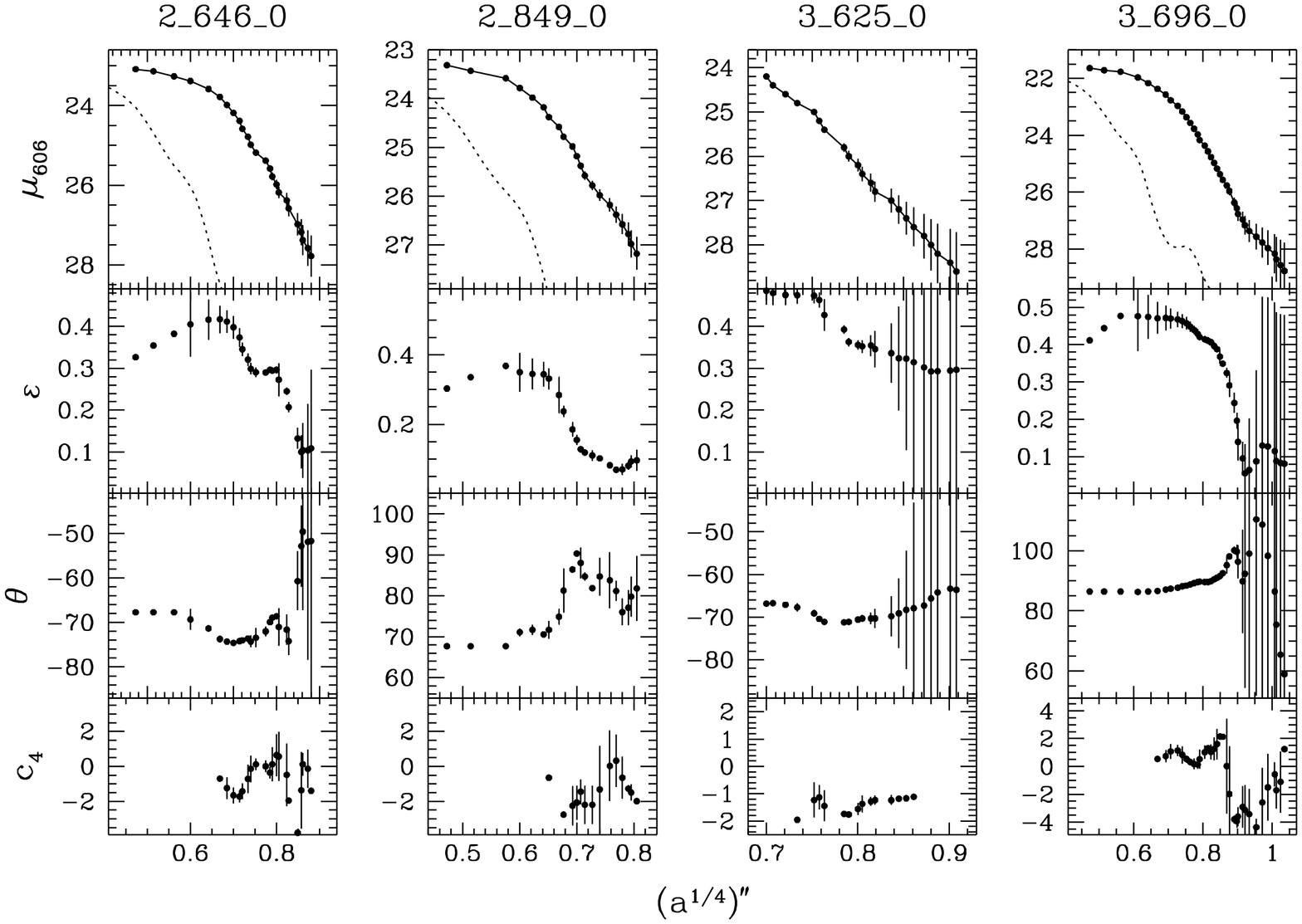,height=11.truecm,width=19.truecm,angle=-90}}
\end{figure*}
\begin{figure*}
\centerline{\psfig{file=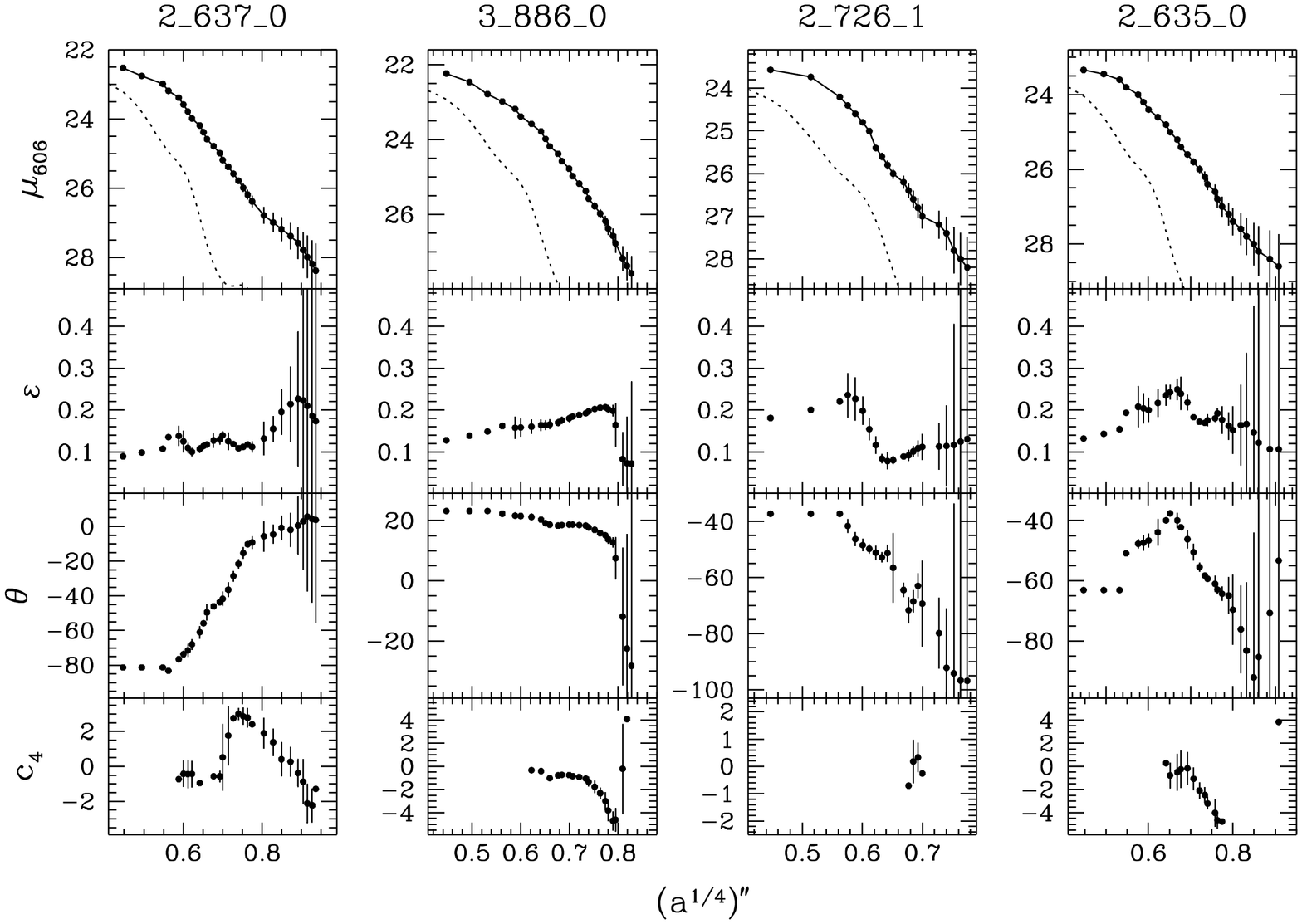,height=11.truecm,width=19.truecm,angle=-90}}
\caption [] {
...continue...}
\end{figure*}

\setcounter{figure}{0}
\begin{figure*}
\centerline{\psfig{file=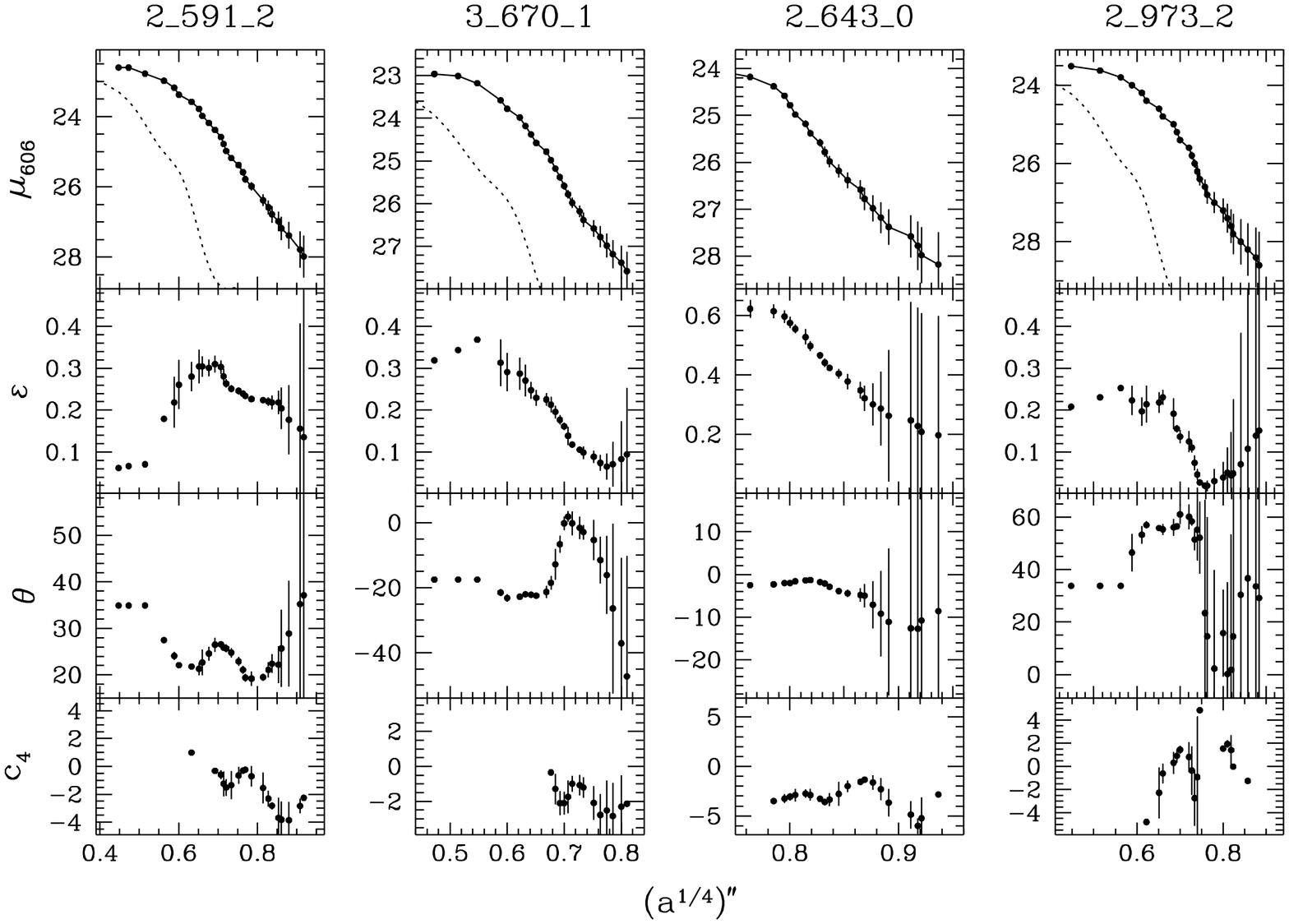,height=11.truecm,width=19.truecm,angle=-90}}
\end{figure*}
\begin{figure*}
\centerline{\psfig{file=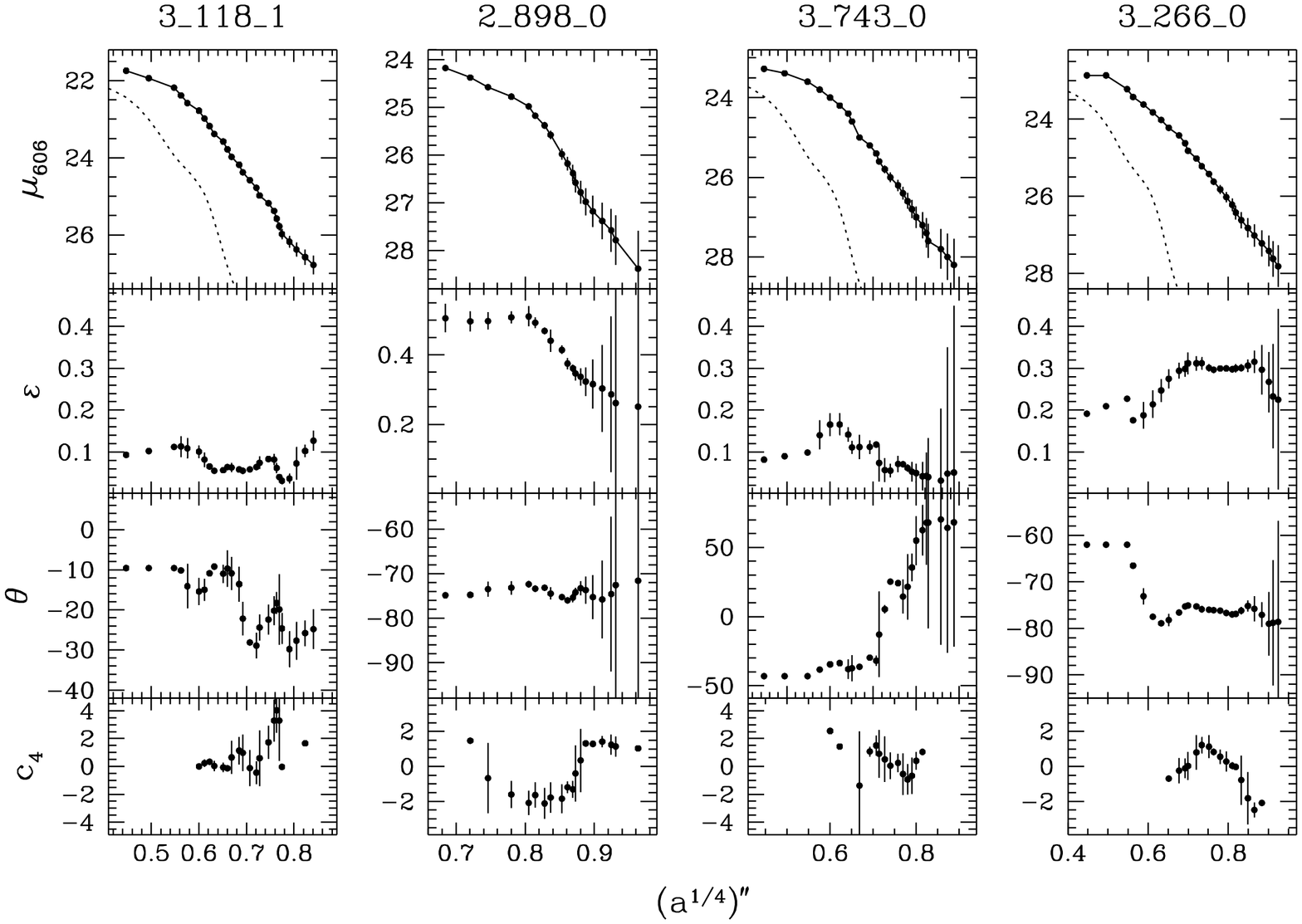,height=11.truecm,width=19.truecm,angle=-90}}
\caption [] {
...continue...}
\end{figure*}

\setcounter{figure}{0}
\begin{figure*}
\centerline{\psfig{file=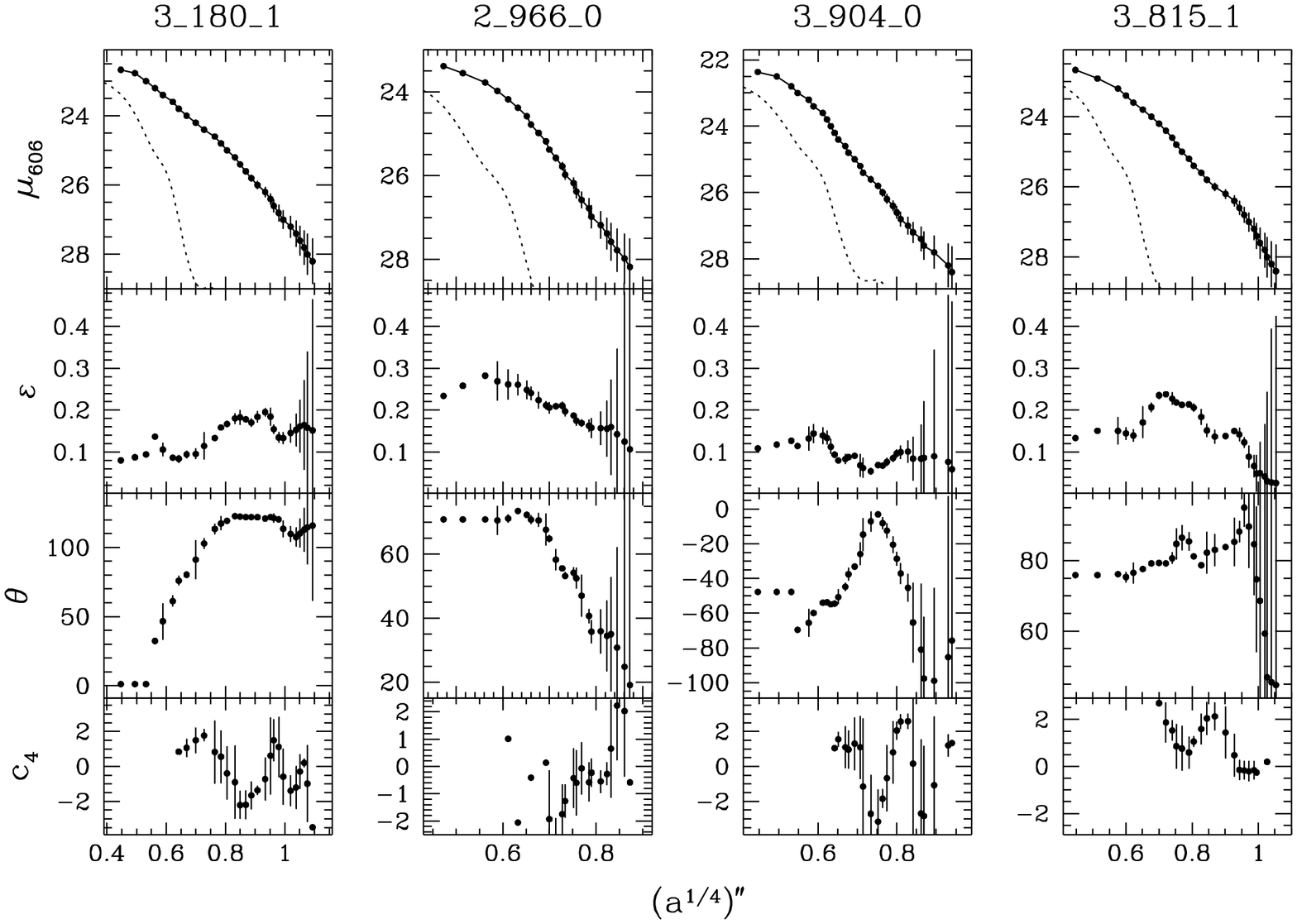,height=11.truecm,width=19.truecm,angle=-90}}
\end{figure*}
\begin{figure*}
\centerline{\psfig{file=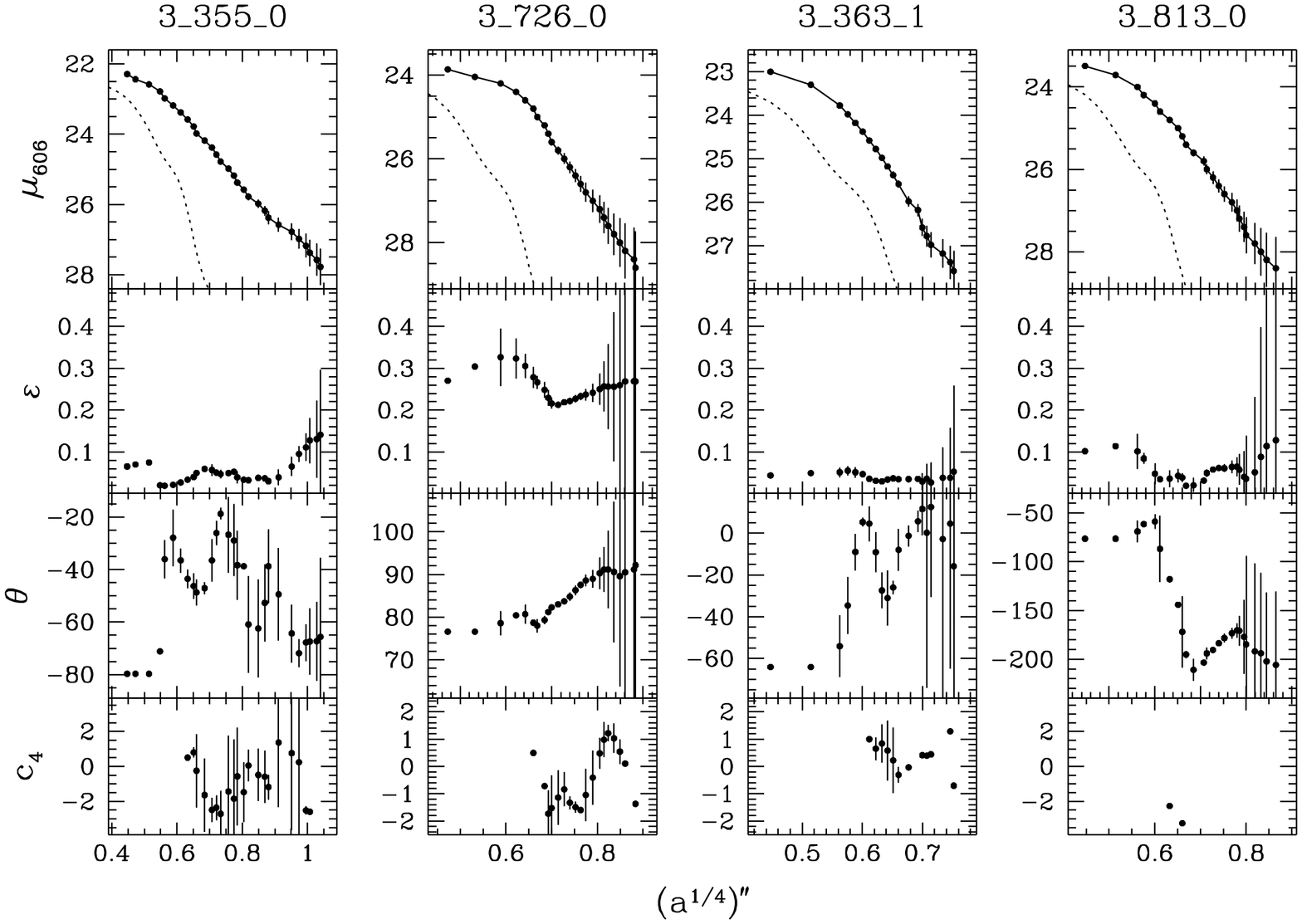,height=11.truecm,width=19.truecm,angle=-90}}
\caption [] {
...continue...}
\end{figure*}

\setcounter{figure}{0}
\begin{figure*}
\centerline{\psfig{file=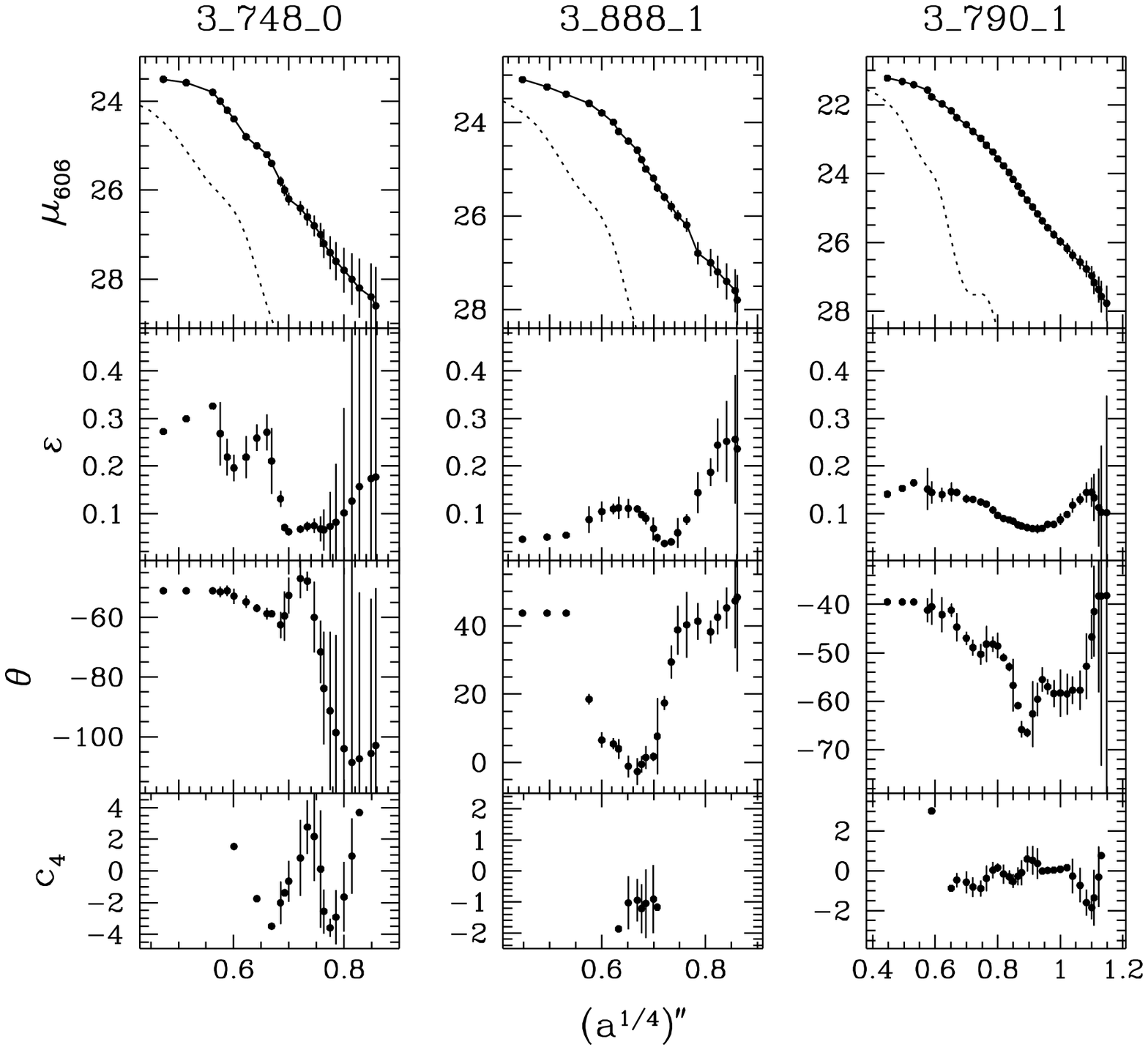,height=11.truecm,width=19.truecm,angle=-90}}
\caption [] {
...continue}
\end{figure*}

\clearpage
\end{document}